\documentclass[aps,prc,twocolumn,showpacs,nofootinbib,aps,letterpaper]{revtex4-1}

\usepackage{amsmath}
\usepackage{amsfonts,amssymb}
\usepackage{multirow}
\usepackage[dvips]{graphicx}
\usepackage{float}
\usepackage{appendix}
\usepackage{color}
\usepackage{url}

\usepackage{subfigure}

\def\beq{\begin{equation}}
\def\eeq{\end{equation}}
\def\bea{\begin{eqnarray}}
\def\eea{\end{eqnarray}}

\def\nl{\nonumber\\}
\def\jpsi{J/\psi}
\def\psip{\psi'}
\def\ppbar{\bar p p}
\def\nnbar{\bar n n}
\def\NNbar{\bar N N}

\RequirePackage{xspace} \allowdisplaybreaks

\begin{document}

\title{On the near-threshold {\boldmath$\ppbar$} invariant mass spectrum measured 
in {\boldmath$\jpsi$} and {\boldmath$\psip$} decays} 
\author{Xian-Wei Kang$^{1}$}\email{x.kang@fz-juelich.de}
\author{Johann Haidenbauer$^1$}\email{j.haidenbauer@fz-juelich.de}
\author{Ulf-G. Mei\ss ner$^{1,2}$}\email{meissner@hiskp.uni-bonn.de}
\affiliation{$^1$Institute for Advanced Simulation, J\"ulich
Center for Hadron Physics, and Institut f\"ur
Kernphysik, Forschungszentrum J\"ulich, D-52425 J\"ulich, Germany\\
$^2$Helmholtz-Institut f\"ur Strahlen- und Kernphysik and Bethe
Center for Theoretical Physics, Universit\"at Bonn, D-53115 Bonn,
Germany}

\begin{abstract}
A systematic analysis of the near-threshold enhancement in the $\ppbar$
invariant mass spectrum seen in the decay reactions
$\jpsi \to x \ppbar$ and $\psip(3686) \to x \ppbar$ $(x = \gamma,\, \omega,\, 
\rho,\, \pi,\, \eta)$ is presented. 
The enhancement is assumed to be due to the $\NNbar$ final-state interaction
(FSI) and the pertinent FSI effects are evaluated in an approach that is 
based on the distorted-wave Born approximation. 
For the $\NNbar$ interaction a recent potential derived within chiral effective
field theory and fitted to results of a partial-wave analysis of 
$\ppbar$ scattering data is considered and, in addition, an older 
phenomenological model constructed by the J\"ulich group. 
It is shown that the near-threshold spectrum observed in various decay reactions
can be reproduced simultaneously and consistently by our treatment of the $\ppbar$ FSI. 
It turns out that the interaction in the isospin-1 $^1S_0$ channel required for the
description of the $\jpsi \to \gamma \ppbar$ decay predicts a $\NNbar$ bound state. 
\end{abstract}

 \maketitle

\section{Introduction}

The origin of the enhancement in the antiproton-proton ($\bar p p$) mass spectrum 
at low invariant masses
observed in heavy meson decays like $J/\psi \to \gamma \ppbar$, 
$B \to K \ppbar$ and $\bar B \to D \ppbar$, but also in the reaction 
$e^+ e^- \leftrightarrow \ppbar$, is an interesting and still controversially 
discussed issue. 
In particular, the spectacular near-threshold enhancement in
the $\ppbar$ invariant mass spectrum for the reaction
$J/\psi \to \gamma \ppbar$, first observed in a high-statistics and
high-mass-resolution experiment by the BES Collaboration~\cite{gamma2003},
has led to numerous publications with speculations about the discovery 
of a new resonance~\cite{gamma2003} or of a $\ppbar$ bound state 
(baryonium)~\cite{Datta,MLYan1,MLYan2}, and was even associated with
exotic glueball states~\cite{Kochelev2005,Li2005,He2005}.
However, in the above processes the hadronic final-state
interaction (FSI) in the $\ppbar$ system should play a role too.
Indeed, the group in J\"ulich-Bonn \cite{SibirtsevPRD,Haidenbauer2} but also others
\cite{Kerbikov,Bugg,Zou:2003zn,Loiseau1,Entem,Loiseau2,Chen,Loiseau3} 
demonstrated that the near-threshold enhancement in the $\ppbar$ invariant mass
spectrum of the reaction $J/\psi \to \gamma {\ppbar}$ could be simply 
due to the FSI between the outgoing proton 
and antiproton. Specifically, the calculation \cite{SibirtsevPRD,Haidenbauer2}
based on the realistic J\"ulich antinucleon--nucleon
($\NNbar$) model~\cite{modelA1,modelA2,modelA3},
the one by the Paris group~\cite{Loiseau2}, utilizing the Paris
$\NNbar$ model \cite{Paris2009}, and that of Entem and Fern\'andez~\cite{Entem},
using a $\NNbar$ interaction derived from a constituent quark model~\cite{Entem06},
explicitly confirmed the significance of FSI effects estimated in the
initial studies~\cite{Kerbikov,Bugg,Zou:2003zn}
within the effective range approximation.

In the present work we perform a systematic analysis of the near threshold 
enhancements in the reactions $\jpsi \to x \ppbar$ and 
$\psip(3686) \to x \ppbar$ $(x = \gamma,\, \omega,\, 
\rho,\, \pi,\, \eta)$
with emphasis on the role played by the $\ppbar$ interaction.
The aim is to achieve a simultaneous and consistent description 
of all $\ppbar$ invariant mass spectra measured in the various
reactions. 
FSI effects for different decay channels cannot be expected to be 
quantitatively the same. In particular, with regard to $\ppbar$, the 
two baryons have to be in different states if the quantum numbers of
the third particle in the decay channel differ, in accordance with 
the general conservation laws. Furthermore, it is possible that dynamical 
selection rules, reflecting the details of the reaction mechanism, 
could suppress the decay into $\ppbar$ $S$-waves for some decays 
near threshold. Thus, in different decay modes the final ${\ppbar}$ 
system can and must be in different partial waves and, accordingly 
the FSI effects will differ too.
 
As mentioned, initial studies of FSI effects in the decay $\jpsi \to \gamma \ppbar$ 
were done in the rather simplistic effective range approximation. Later 
investigations, like the ones performed by us \cite{SibirtsevPRD,Haidenbauer2},
employed directly scattering amplitudes from realistic $\NNbar$ potential 
models. Still also here the treatment of the FSI is done within the so-called 
Migdal-Watson approach \cite{Watson,Migdal} where the elementary decay (or 
production) amplitude is simply multiplied with the $\ppbar$ $T$-matrix.
It is known that this approach works reasonably well for reactions with a 
final $NN$ system \cite{HanhartR}. 
In this case the scattering length $a$ is fairly large, 
for example, $a \approx -24$~fm for a final $np$ system (in the $^1S_0$ state).
Measurements of the level shifts in antiprotonic hydrogen atoms suggest 
that the scattering lengths for $\ppbar$ scattering are presumably only 
in the order of $1$ to $2$~fm \cite{Gotta}. Moreover, those 
scattering lengths are complex due to the presence of annihilation 
channels. Therefore, in the present paper we consider an alternative 
and more refined approach for taking into account the FSI. 
Specifically, we use the Jost function which is calculated directly 
from realistic $\NNbar$ potentials. FSI effects are then taken 
into account by multiplying the reaction amplitude with the inverse
of this Jost function. This is practically equivalent to a treatment 
of such decay reactions within a distorted-wave Born approximation.
Note that this is different from the popular Jost-function 
approach based on the effective range approximation \cite{Watsonbook}
which is widely used in investigations of FSI effects.

We present  results for the decays 
$\jpsi \to x \ppbar$ with $x = \gamma, \ \omega, \ \pi^0, \ \eta$,
which all have been measured. 
For the last three cases parity, $G$-parity, and isospin are
conserved so that each of those channels allows one to explore
the ${\ppbar}$ system in a distinct partial wave.
At the same time the analoguous reactions $\psip \to x \ppbar$ are 
studied. In this case there are data for $x = \gamma,\, \pi^0,\, \eta$.
Clearly, if ${\ppbar}$ FSI effects are responsible for the enhancements 
seen in specific $\jpsi$ decays, then very similar effects should occur 
in the corresponding $\psip$ decays because the selection rules are the 
same.

As far as the $\NNbar$ interaction is concerned we employ again the 
phenomenological model A(OBE) of the J\"ulich group \cite{modelA1} used
in our earlier works 
\cite{SibirtsevPRD,Haidenbauer1,Haidenbauer2,Haidenbauer3}.
In addition, and as a novelty, we utilize also a $\NNbar$ interaction
derived in the framework of chiral effective field theory (EFT) \cite{JHEP}. 
The latter interaction incorporates results of a recent partial-wave analysis (PWA) 
of $\ppbar$ scattering data \cite{Zhou}. In particular, this EFT potential has 
been constructed in such a way, that it reproduces the amplitudes determined in 
the PWA well up to laboratory energies of $T_{lab} \approx 200-250$ MeV \cite{JHEP},
i.e. in the low-energy region where we expect that FSI effects are important.

As pointed out at the beginning, also in decays of the $B$ and $\Upsilon$ mesons 
to final states with a $\NNbar$ pair enhancements at low invariant masses have been 
observed \cite{Belle1,Belle2,Belle5,Aubert06,delAmo11,Aaij13,Aaij14,Athar05}. 
However, in the majority of those experiments the invariant-mass resolution of 
the $\NNbar$ spectrum is relatively low and often there are only two or three data 
points in the (relevant) near-threshold region. Therefore, we refrain from looking at 
those data in detail.
Note also, that in case of weak decays like $B \to K\ppbar$ or $B \to D\ppbar$ parity is not
conserved and, as a consequence, there is less restriction on the possible partial waves
of the $\NNbar$ final state. 
The situation is different for the reaction $e^+ e^- \leftrightarrow \ppbar$. 
As shown by us in recent studies~\cite{Haidenbauere1,Haidenbauere2}, employing the same 
formalism and the same $\NNbar$ interactions as in the present work,
the FSI mechanism can indeed explain the near-threshold enhancement seen in 
the data taken by the PS170 \cite{Bardin}, the FENICE \cite{Antonelli} and the 
BaBar \cite{Lees} Collaborations.

The paper is structured in the following way: 
In Section II we provide a summary of the formalism that we employ for treating the
FSI due to the $\NNbar$ interaction. We discuss also the selection
rules for the decay channels considered.
Results of our calculations are presented in Section III. First we analyze 
hadronic decay channels of $\jpsi$ and $\psip$ (where isospin is assumed to be conserved)
and compare our predictions with measurements of the $\ppbar$ invariant mass spectrum
for the $\pi^0\ppbar$, $\eta\ppbar$, and $\omega\ppbar$ channels. 
Subsequently we consider radiative decays. Since it turns out that the $\ppbar$ 
invariant mass spectrum of $\jpsi \to \gamma\ppbar$ can no longer be described with 
the employed and previously established $\NNbar$ interactions, once the 
more realistic treatment of FSI effects is utilized, we perform and 
present a refit of the chiral EFT $\NNbar$ potential that reproduces the $\gamma\ppbar$ data
and stays also very close to the result of the PWA (and to the original EFT potential \cite{JHEP}) 
for the relevant ($^1S_0$) partial wave. 
The paper ends with a summary.
Results of the refitted $^1S_0$ $\NNbar$ potential are presented in an appendix, and compared
with the PWA and the previously published EFT potential \cite{JHEP}.

\section{Treatment of the $\NNbar$ final state interaction}
 
Our study of the processes of $\jpsi$ (or $\psip$) decaying to 
$x\ppbar$ ($x=\gamma,\,\omega,\,\pi,\,\eta$) is based on the distorted wave Born approximation (DWBA)
where the reaction amplitude $A$ is given by 
\begin{equation}
A=A^0 + A^0 G^{\NNbar} T^{\NNbar} \ . 
\label{eq:DWBA}
\end{equation}
Here $A^0$ is the elementary (or primary) decay amplitude, $G^{\NNbar}$ the free $\NNbar$ Green's 
function, and $T^{\NNbar}$ the $\NNbar$ scattering amplitude. 
For a particular (uncoupled) $\NNbar$ partial wave with orbital angular momentum $L$, Eq.~(\ref{eq:DWBA}) reads
\begin{equation}\label{eq:fullFSI}
A_L=A_L^{0}+\int_0^\infty \frac{dp p^2}{(2\pi)^3} A_L^{0} 
\frac{1}{2E_k-2E_p+i0^+}T_{L}(p,k;E_k) ,
\end{equation}
where $T_{L}$ denotes the partial-wave projected $T$-matrix element, and $k$ and $E_k$ are
the momentum and energy of the proton (or antiproton) in the center-of-mass system of the $\NNbar$ pair.
The quantity $T_{L}(p,k;E_k)$ is obtained from the solution of the Lippmann-Schwinger (LS) equation,
\begin{eqnarray}
&&T_{L}(p',k;E_k)=V_{L}(p',k)+ \nl
&& \int_0^\infty \frac{dpp^2}{(2\pi)^3} \, V_{L}(p',p)
\frac{1}{2E_k-2E_p+i0^+}T_{L}(p,k;E_k)~,\nl 
\label{LS}
\end{eqnarray}
for a specific $\NNbar$ potential $V_L$. In case of coupled partial waves like the $^3S_1$--$^3D_1$ 
we solve the corresponding coupled
LS equation as given in Eq.~(2.20) of Ref.~\cite{JHEP}, and use then $T_{LL}$ in Eq.~(\ref{eq:fullFSI}). 
 
In principle, the elementary production amplitude $A^0_L$ in Eq.~\eqref{eq:fullFSI} has
an energy dependence and it depends also on the $\NNbar$ momentum and the photon momentum
relative to the $\NNbar$ system. However, in the near-threshold region the variation of the 
production amplitude with regard to those variables should be rather small as compared to the strong
momentum dependence induced by the $\NNbar$ FSI and, therefore, we neglect
it in the following. Then Eq.~\eqref{eq:fullFSI} can be reduced to 
\begin{widetext}
\begin{equation}\label{eq:Jost}
A_L=\bar A^0_L k^L\left[1+ \int_0^\infty \frac{dp p^2}{(2\pi)^3} 
\frac{p^L}{k^L}\frac{1}{2E_k-2E_p+i0^+}T_{L}(p,k;E_k)\right]
=\bar A^0_L k^L \psi_{k,L}^{(-)*}(0) \ . 
\end{equation}
\end{widetext}
Here, we have separated the factor $k^L$ which ensures the correct threshold behaviour 
for a particular orbital angular momentum so that $\bar A^0_L$ is then a constant. 
The quantity in the bracket in Eq.~\eqref{eq:Jost} is the so-called enhancement 
factor \cite{Watsonbook}. Introducting a suitably normalized wave function for
the $\ppbar$ pair in the continuum \cite{Watsonbook}, $\psi_{k,L}^{(-)*}(0)$, this 
quantity is 
just the inverse of the Jost function, i.e. $\psi_{k,L}^{(-)*}(0)=f_L^{-1}(-k)$. 
We want to emphasize that in the present work we calculate the enhancement factor 
for the considered $\NNbar$ interactions explicitly, which amounts to an integral over 
the pertinent (half-off-shell) $T$ matrix elements, see Eq.~(\ref{eq:Jost}).
This should not be confused with the popular Jost-function approach which relies
simply on the effective range approximation. In any case, the latter cannot be easily
applied in the $\NNbar$ case because now the scattering length as well as the
effective range are complex quantities. 
For a thorough discussion of various aspects of the treatment of FSI effects due
to baryon-baryon interactions, see Refs.~\cite{Hanhart,Baru,Gasparyan}.

The differential decay rate for the processes  $X \to x\ppbar$ ($X = \jpsi, \ \psip$) 
can be written in the form \cite{SibirtsevPRD,kinematicbook}
\beq \label{eq:decayrate}
\frac{d\Gamma}{dM}=\frac{\lambda^{1/2}(m_{X}^2,M^2,m_x^2)\sqrt{M^2-4m_p^2}}{2^6\pi^3m_{X}^2}|A|^2 \ ,
\eeq
after integrating over the angles.
Here the K\"all\'en function $\lambda$ is defined as $\lambda(x,y,z)=\left((x-y-z)^2-4yz\right)/(4x)$, 
$M\equiv M(\ppbar)$ is the invariant mass of the $\ppbar$ system, $m_{X}$, $m_p$, $m_x$ are the 
masses of the $\jpsi$ (or $\psip$), the proton, and the meson (or $\gamma$) in the final state, 
while $A$ is the total (dimensionless) reaction amplitude.
Note that in Eq.~\eqref{eq:decayrate} we have assumed that averaging over the spin states has been
already performed \cite{kinematicbook}. In the present manuscript we will consider only individual 
partial wave amplitudes and, therefore, use a specific $A_L$ in Eq.~(\ref{eq:decayrate}).

Let us come back to $A^0$ and, specifically, to the assumption that it is constant in the 
region near the $\NNbar$ threshold where we perform our calculation. Such an assumption is 
sensible if there are no dominant one-, two- or even three-particle doorway channels,
with masses or thresholds close to the $\NNbar$ threshold, for the transition from $\jpsi$ 
(or $\psip$) to $\NNbar$.
For example, a dominant $\NNbar$ production via $\rho$, $\pi\pi$ or $\pi\pi\pi$ intermediate
states would definitely not invalidate this assumption. However, a genuine resonance
with a mass comparable to the $X(1835)$ found by the BES Collaboration in the reaction 
$\jpsi\to\gamma\pi^+\pi^-\eta'$ \cite{Ablikim:X1835,Ablikim:X1835a} would render it already 
somewhat questionable, if it constitutes indeed the dominant doorway channel for the decay 
into the $\NNbar$ system. 
In any case, and as in all previous 
works that exploit FSI effects, it should be clear that the assumption of a constant $\bar A^0_L$
is first and foremost a working hypothesis. The question that can be addressed in our study is 
simply, whether the energy dependence generated by the $\NNbar$ interaction in the final 
state alone suffices to describe the $\ppbar$ invariant mass spectra or not. A possible 
genuine energy dependence of the primary production amplitude itself cannot be excluded.  

\begin{table}[htbp]
\begin{center}
\begin{tabular}{ccc}\hline\hline
channels  & partial waves  & isospin\\
\hline \hline
$\jpsi\to\gamma\ppbar$  &\multirow{2}{*}{$^1S_0 \, [0^{-+}], {}^3P_0 \, [0^{++}], {}^3P_1 \, [1^{++}], 
{}^3P_2 \, [2^{++}]$}  &\multirow{2}{*}{0, 1}\\
$\psip\to\gamma\ppbar$  &   &   \\
\hline
$\jpsi\to\omega\ppbar$ &\multirow{2}{*}{$^1S_0, {}^3P_0, {}^3P_1, {}^3P_2$}  & \multirow{2}{*}{0}\\
$\psip\to\omega\ppbar$  &   &   \\
\hline
$\jpsi\to\rho\ppbar$ & \multirow{2}{*}{$^1S_0, {}^3P_0, {}^3P_1, {}^3P_2$}  & \multirow{2}{*}{1}\\
$\psip\to\rho\ppbar$  &   &   \\
\hline
$\jpsi\to\eta\ppbar$ & \multirow{2}{*}{$^3S_1 \, [1^{--}], {}^1P_1 \, [1^{+-}]$}  & \multirow{2}{*}{0}\\
$\psip\to\eta\ppbar$ &   &  \\
\hline
$\jpsi\to\pi^0\ppbar$ & \multirow{2}{*}{$^3S_1, {}^1P_1$}  & \multirow{2}{*}{1}\\
$\psip\to\pi^0\ppbar$ &   &  \\
\hline
$\chi_{c0}\to\pi^-\bar n p$ & $^1S_0, {}^3P_1$  & 1\\
 \hline\hline
\end{tabular}
\caption{Allowed $\NNbar$ partial waves, $J^{PC}$ assignments and isospins 
for various channels up to $P$ waves.}
\label{tab:JPC}
\end{center}
\end{table}

Conservation of the total angular momentum, together with parity, charge conjugation 
and isospin conservation for the strong interactions, put strong constraints on the 
partial waves of the produced $\ppbar$ system. 
We list the allowed quantum numbers for various decay channels in Table~\ref{tab:JPC} 
for orbital angular momentum $L \le 1$, i.e. $S$ and $P$ waves. 
We use the standard notation $^{(2S+1)}L_J$, where $L, S, J$ are the orbital angular momentum, the
total spin and the total angular momentum. The isospin $I$ is sometimes indicated by 
the notation $^{(2I+1)(2S+1)}L_J$. 
In the actual calculation we consider, in general, only the lowest partial wave, i.e. 
either the $^1S_0$ or the $^3S_1$. 
Those should be the dominant partial waves for energies near the $\ppbar$ threshold.
As already said, we assume also that a single partial wave saturates (or dominates) in 
the energy range covered, i.e. up to excess energies of 
$M(\ppbar)-2m_p\approx 100\,\text{MeV}$ considered also in the earlier works 
\cite{SibirtsevPRD,Haidenbauer1,Haidenbauer2,Haidenbauer3,Loiseau1,Entem,Loiseau2}. 
In principle, higher partial wave may well play a non-negligible role around 100 MeV (or
even at somewhat lower energies) and one could limit oneself to excess energies up to 
$\approx 50$ MeV,
say, to be on the safe side. Or one could introduce a cocktail of amplitudes. 
However, at present there is very little experimental information to constrain the 
relative weight of the partial waves and also their interference. Hopefully in the 
future, with a larger data set and more precise measurements of angular distributions
a more refined analysis will become feasible. 
Note that it is possible that dynamical selection rules lead to a suppression 
of the lowest partial waves in the $\NNbar$ system. This could be also detected by 
measuring the angular distributions of the decay products.

\section{Results}

Most of the studies of FSI effects in the reaction $\jpsi\to\gamma\ppbar$ 
(and related decays) in the literature are performed in the Migdal-Watson 
approach \cite{Watson,Migdal}. 
In this approximation, instead of evaluating the integral equation 
that arises in the DWBA, see Eq.~(\ref{eq:fullFSI}), the FSI is simply accounted 
for by multiplying the elementary reaction amplitude by the on-shell $\NNbar$ 
$T$-matrix, i.e. $A_L \approx N \cdot \bar A^0_L T_L(k,k;E_k)/k_L$, where 
$N$ is an arbitrary normalization factor. It is known from pertinent
studies that the applicability of the Migdal-Watson approach 
is limited to a fairly small energy range \cite{Baru}. In particular, it
works only reasonably well if the scattering length is rather large --
which is the case for $NN$ scattering with values of $a\approx -24$ fm
for the interaction in the ($np)$ $^1S_0$ partial wave. However, for $\NNbar$
scattering the values for the scattering lengths are typcially in the
order of only 1--2 fm \cite{Gotta,JHEP}. 
 
\begin{figure}[htbp]
\begin{center}
\includegraphics[height=60mm,clip]{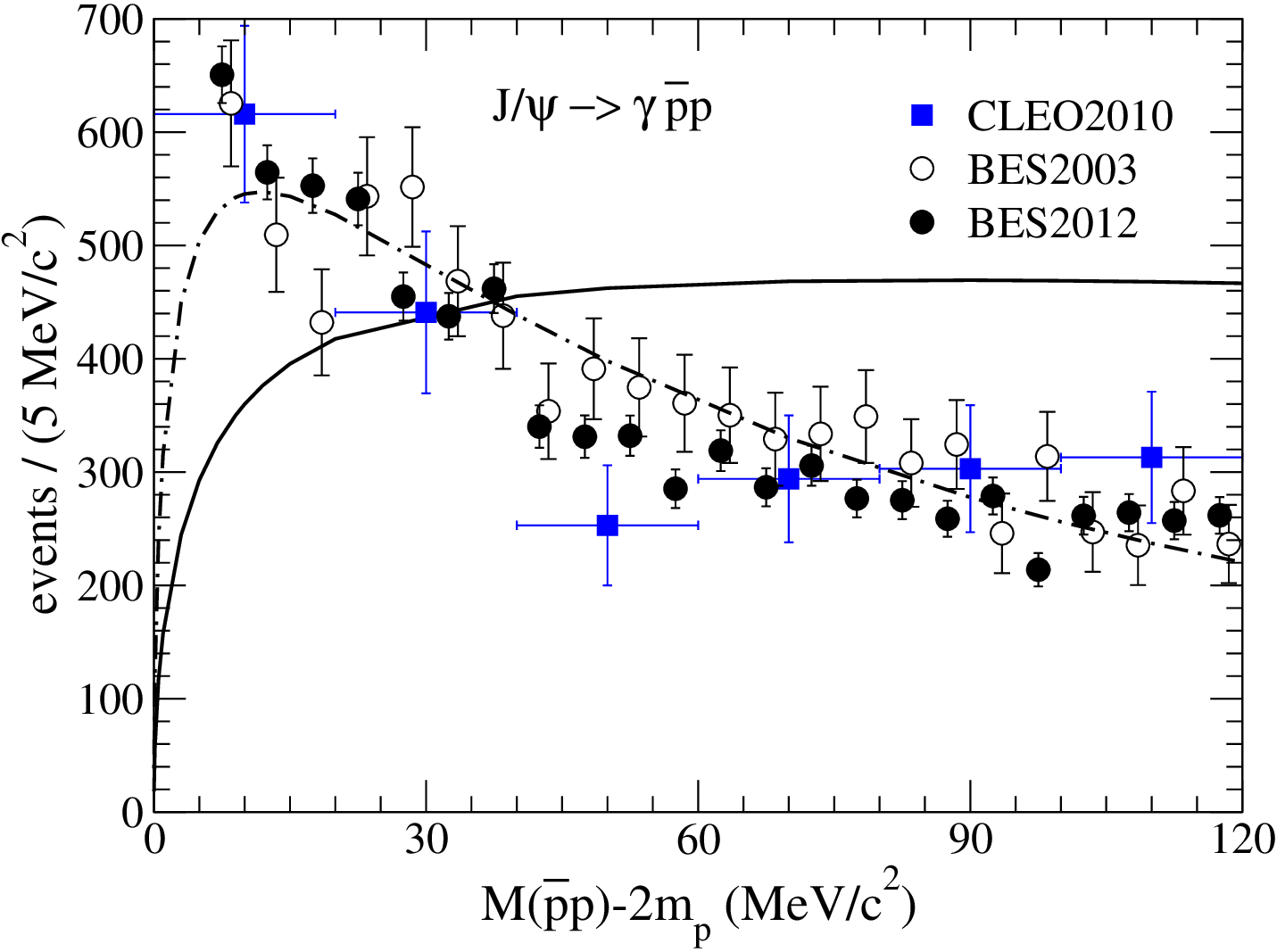}
\caption{$\ppbar$ spectrum for the decay $\jpsi\to\gamma\ppbar$. The solid curve denotes
results for the $\NNbar$ interaction A(OBE) based on the DWBA, see Eq.~(\ref{eq:Jost}),
while the dash-dotted curve is based on the Migdal-Watson approximation \cite{SibirtsevPRD}. 
Data are taken from Refs.~\cite{gamma2003,gamma2012,CLEOpsip}.
The measurement of Ref.~\cite{gamma2012} is adopted for the scale. 
The data for the BES measurement from 2003 have been shifted slightly to the right, cf. text.
}
\label{fig:MW}
\end{center}
\end{figure}

Entem and Fern\'andez have presented results based on the
Migdal-Watson approximation and on the DWBA~\cite{Entem} 
and those suggest drastic differences between the two approaches. 
Indeed, we can confirm this with our own calculation employing the 
$\NNbar$ potential A(OBE) \cite{modelA1} that we used in our earlier studies 
\cite{SibirtsevPRD,Haidenbauer1,Haidenbauer2,Haidenbauer3}. 
Corresponding results are presented in Fig.~\ref{fig:MW}. The dash-dotted curve 
is the prediction for the $\ppbar$ invariant mass based on the
$I=1$ $^1S_0$ amplitude in the Migdal-Watson approximation, as published
in Ref.~\cite{SibirtsevPRD} which reproduces rather well the energy dependence
found in the experiments \cite{gamma2003,gamma2012,CLEOpsip}.
The result for the same $\NNbar$ interaction but based on the more
refined treatment of the FSI, Eq.~(\ref{eq:Jost}), is no longer in agreement 
with the data, see the solid curve in Fig.~\ref{fig:MW}.
At first sight, this is certainly disturbing. However, we want to emphasize 
that it would be premature to see the observed discrepancy as signal for the 
failure of the FSI interpretation of the enhancement in the near-threshold 
$\ppbar$ invariant mass spectrum. Rather it could be simply an evidence for 
certain shortcomings of the employed $\NNbar$ interaction in the
$^1S_0$ partial wave. In addition, 
isospin is not conserved in the reaction $\jpsi\to\gamma\ppbar$ and,
therefore, the actual $\NNbar$ FSI can involve any combination 
of the $I=0$ and $I=1$ $^1S_0$ amplitudes. 
We will address these issues in detail later in this section. First we want to
look at purely hadronic ($\jpsi$ and $\psip$) decay channels 
with a $\ppbar$ final state where nominally isospin is conserved. 

But before that we would like to comment on the normalization. 
Usually only event rates are given for the various experiments. These differ for
different experiments and also for different invariant-mass resolutions. For
the figures presented below, in general, we fix the scale according to the 
experiment with the highest resolution. 
Which data set is used to fix the scale will be emphasized in the pertinent caption. 
The data (and error bars) from other experiments with lower resolution are then 
renormalized to this scale, guided by the eye. Also our theory results are 
renormalized to this scale (guided by the eye) by an appropriate choice of 
$\bar A^0_L$ in Eq.~(\ref{eq:Jost}). The only exception is the $\jpsi\to\gamma\ppbar$
reaction, where the constant $\bar A^0_L$ is fixed via a fit to the $\ppbar$ 
invariant-mass spectrum. 
Note also that the actual values of (most of) the data presented in the various figures 
were not directly available to us. We use here values obtained from digitizing the 
figures of the original publications. 
Finally, for some decays the BES Collaboration has published data sets with different 
statistics but with the same momentum resolution. Since we wanted to include both sets 
in the same figure we shifted the ones from the earlier measurement slightly to the 
right (by $1$\,MeV) so that one can distinguish the two data sets easier in the figure. 
This concerns the $\gamma \ppbar$ and the $\omega \ppbar$ channels.

\subsection{Decays into three hadrons}

Besides $\jpsi\to\gamma\ppbar$ there is also experimental information on $\jpsi$ and $\psip$ decays
into three-body channels involving a $\ppbar$ pair and a pseudo-scalar 
($\pi$, $\eta$) \cite{gamma2003,BESJpsipi,BESpsippi,CLEOpsip,BESJpsieta,BESpsipeta} 
or vector ($\omega$) \cite{BESJpsiomega2008,BESJpsiomega2013}
meson. There is, however, a strong variation in the quality of the data. While in case of
$\jpsi\to\pi^0\ppbar$ and $\jpsi\to\omega\ppbar$ the momentum resolution is excellent and 
comparable to the one for $\jpsi\to\gamma\ppbar$, the bin widths for the other 
reactions are much larger.

\begin{figure}[htbp]
\begin{center}
\includegraphics[height=60mm,clip]{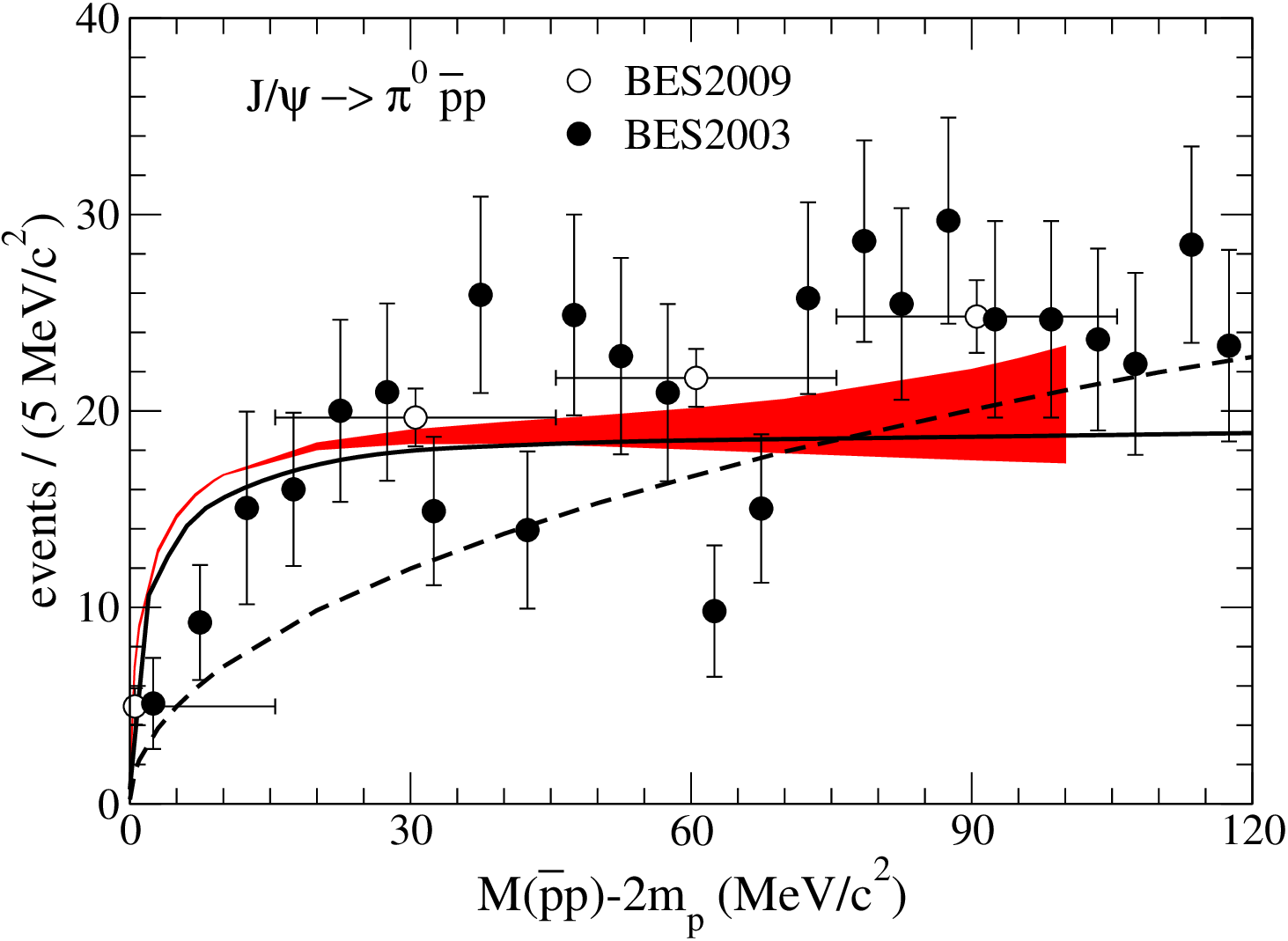}
\caption{$\ppbar$ spectrum for the decay $\jpsi\to\pi^0\ppbar$. 
The band represents the result based on the $\NNbar$ FSI generated from the 
chiral EFT potential \cite{JHEP} while 
the solid curve is the result for the $\NNbar$ interaction A(OBE) \cite{modelA1}.  
The dashed curve denotes the phase space behavior. 
Data are taken from Refs.~\cite{gamma2003,BESJpsipi}.
The measurement of Ref.~\cite{gamma2003} is adopted for the scale. 
}
\label{fig:Jpsipi}
\end{center}
\end{figure}

\begin{figure}[htbp]
\begin{center}
\includegraphics[height=60mm,clip]{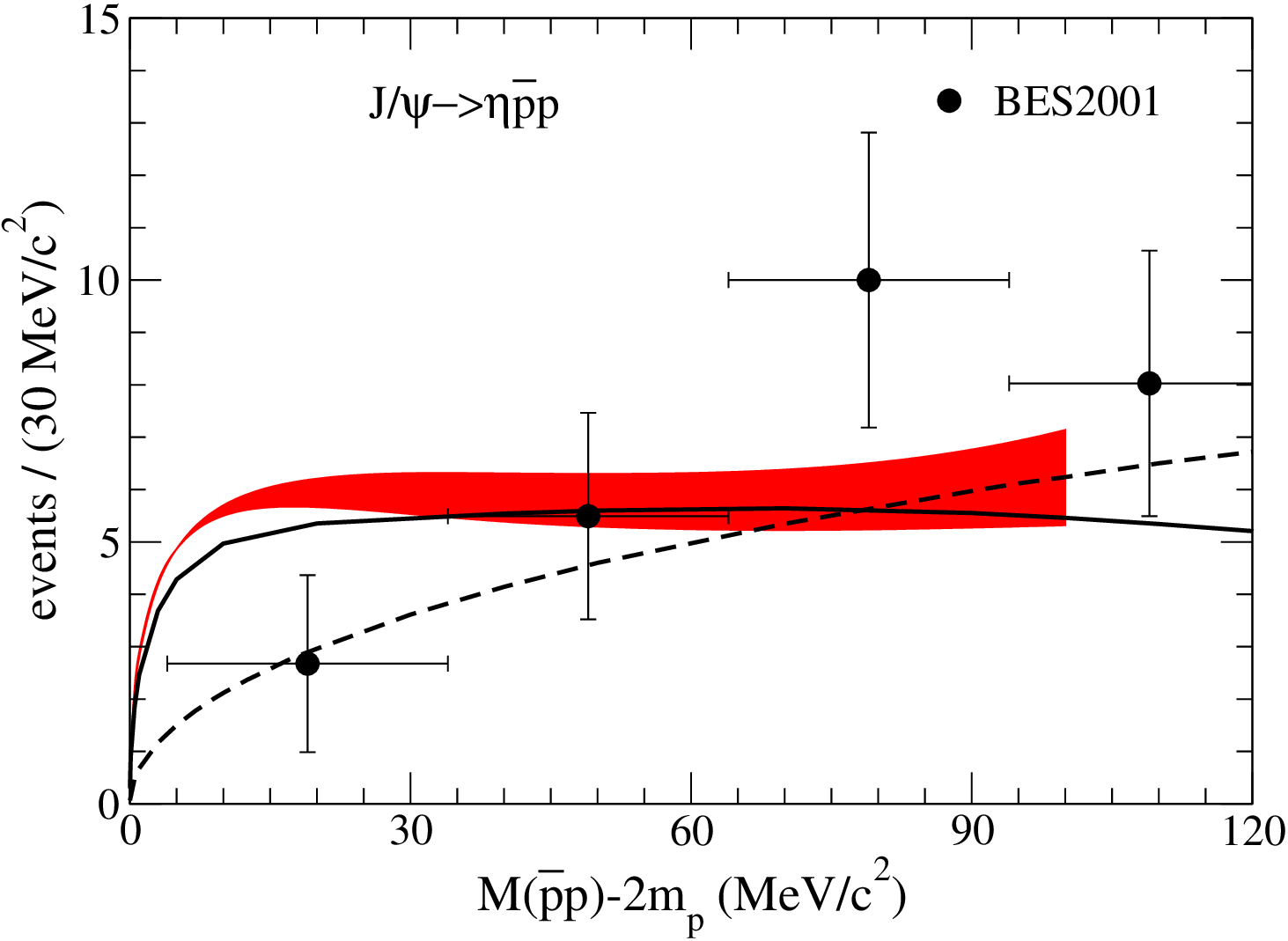}
\caption{$\ppbar$ spectrum for the decay $\jpsi\to\eta\ppbar$.
Same description of curves as in Fig.~\ref{fig:Jpsipi}. 
Data are taken from Ref.~\cite{BESJpsieta}.}
\label{fig:Jpsieta}
\end{center}
\end{figure}

\begin{figure}[htbp]
\begin{center}
\includegraphics[height=60mm,clip]{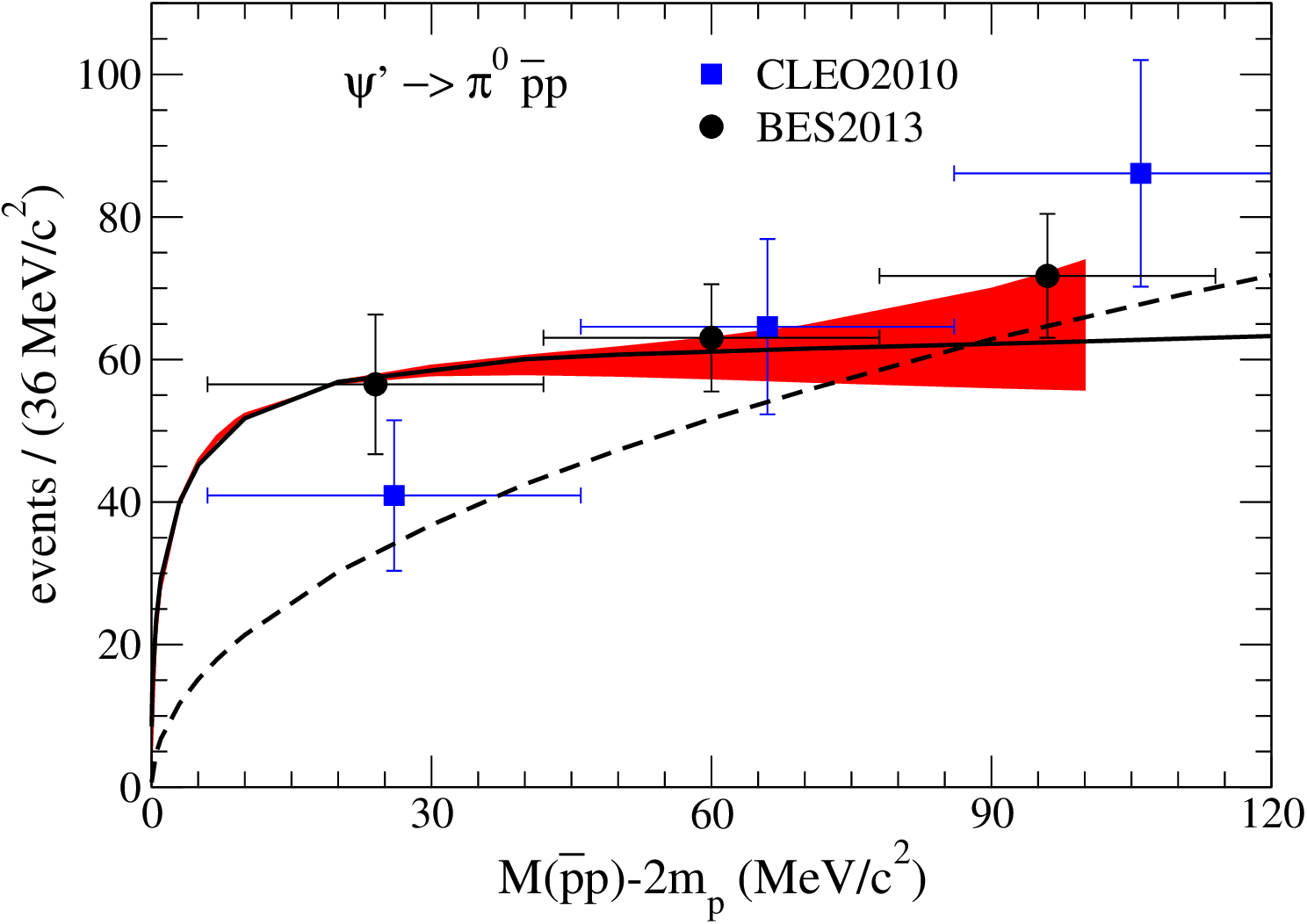}
\caption{$\ppbar$ spectrum for the decay $\psip\to\pi^0\ppbar$.
Same description of curves as in Fig.~\ref{fig:Jpsipi}. 
Data are taken from Refs.~\cite{BESpsippi,CLEOpsip}.
The measurement of Ref.~\cite{BESpsippi} is adopted for the scale. 
}
\label{fig:psippi}
\end{center}
\end{figure}

\begin{figure}[htbp]
\begin{center}
\includegraphics[height=60mm,clip]{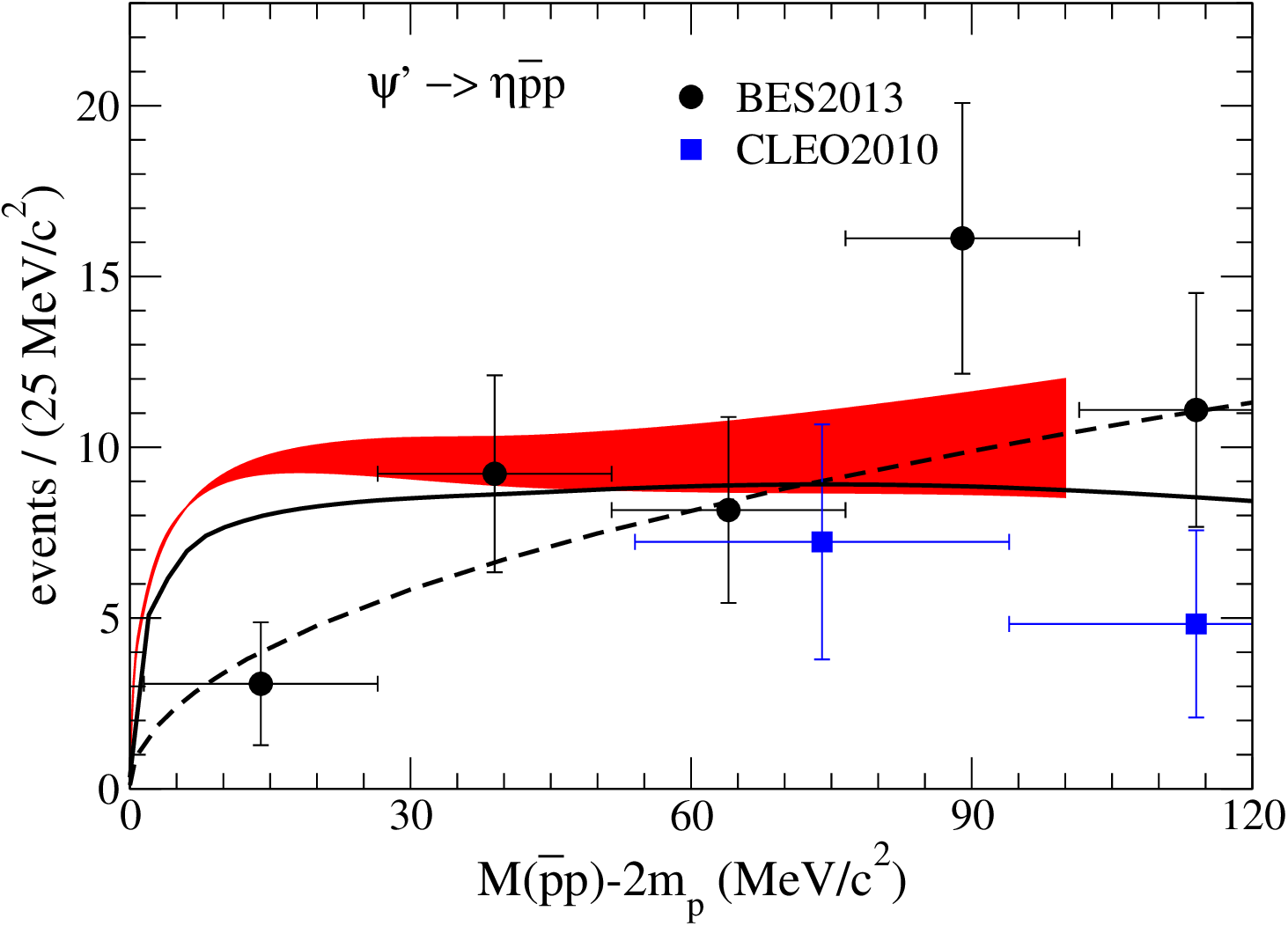}
\caption{$\ppbar$ spectrum for the decay $\psip\to\eta\ppbar$.
Same description of curves as in Fig.~\ref{fig:Jpsipi}. 
Data are taken from Refs.~\cite{BESpsipeta,CLEOpsip}.
The measurement of Ref.~\cite{BESpsipeta} is adopted for the scale. 
}
\label{fig:psipeta}
\end{center}
\end{figure}

Let us first consider channels with pseudo-scalar mesons. 
The processes of $\jpsi$ and $\psip$ decaying to $\pi\ppbar$ or $\eta\ppbar$ involve the $^3S_1$ partial 
wave, see Table~\ref{tab:JPC}. 
The event rates calculated via Eqs.~(\ref{eq:Jost}) and (\ref{eq:decayrate}) are shown in 
Fig.~\ref{fig:Jpsipi} for the decay $\jpsi\to\pi^0\ppbar$, in Fig.~\ref{fig:Jpsieta} for $\jpsi\to\eta\ppbar$,
in Fig.~\ref{fig:psippi} for $\psip\to\pi^0\ppbar$, and in Fig.~\ref{fig:psipeta} for $\psip\to\eta\ppbar$. 
Results for our $\NNbar$ potential derived in chiral EFT are presented as bands.
This band is generated from the four cutoff combinations $\{\Lambda\, , \tilde\Lambda\}$ considered in 
the construction of the EFT $\NNbar$ potential \cite{JHEP} and can be viewed as a (rough) estimate of 
the theoretical uncertainty, see the corresponding discussions in Refs.~\cite{JHEP,SFR}. 
The solid line is the prediction for the meson-exchange potential A(OBE).
The dashed line represents the phase space behavior and follows from Eq.~(\ref{eq:decayrate}) 
by setting the production amplitude $A$ to a constant. In general, the latter is normalized in such 
a way that it coincides with the results for the EFT interaction for excess energies around $70-80$~MeV. 
We want to stress once more that in Fig.~\ref{fig:Jpsipi} and in the other figures in this section
all normalizations are arbitrary. We are only interested in the energy dependence as it 
follows from the FSI effects predicted by the employed $\NNbar$ interactions. 
 
Obviously, in all cases our predictions are in line with the data. Specifically, the 
results for $\jpsi\to\pi^0\ppbar$ are in nice agreement with the experiment. Here the FSI
generates a moderate but noticable enhancement at small $\ppbar$ invariant masses as compared 
to the phase space and yields a $\ppbar$ spectrum which is seemingly closer to the trend 
exhibited by the data than the phase-space curve. It is interesting to see that the results 
based on the chiral potential and on A(OBE) are fairly similar. In this context 
let us remind the reader that we had to introduce some phenomenological adjustments in our earlier 
study based on the Migdal-Watson approximation (and with A(OBE)) in order to be able to 
reproduce that experimental invariant mass spectrum, cf. Eq. (8) in Ref.~\cite{SibirtsevPRD}. 
Now the behavior follows directly from the refined treatment of FSI effects via Eq.~(\ref{eq:Jost}). 
 
The results for the other channels are less conclusive. The invariant-mass resolution in the 
pertinent measurements is only in the order of $30$~MeV or so and, 
consequently, there are only three or four data points below the excess energy of $100$~MeV. 
Whether or not the present data require the enhancement provided by the $\NNbar$ FSI is
difficult to judge. Hopefully, future measurements with much higher statistics as well 
as much higher resolution will provide a more serious test for FSI effects. 

\begin{figure}[htbp]
\begin{center}
\includegraphics[height=60mm,clip]{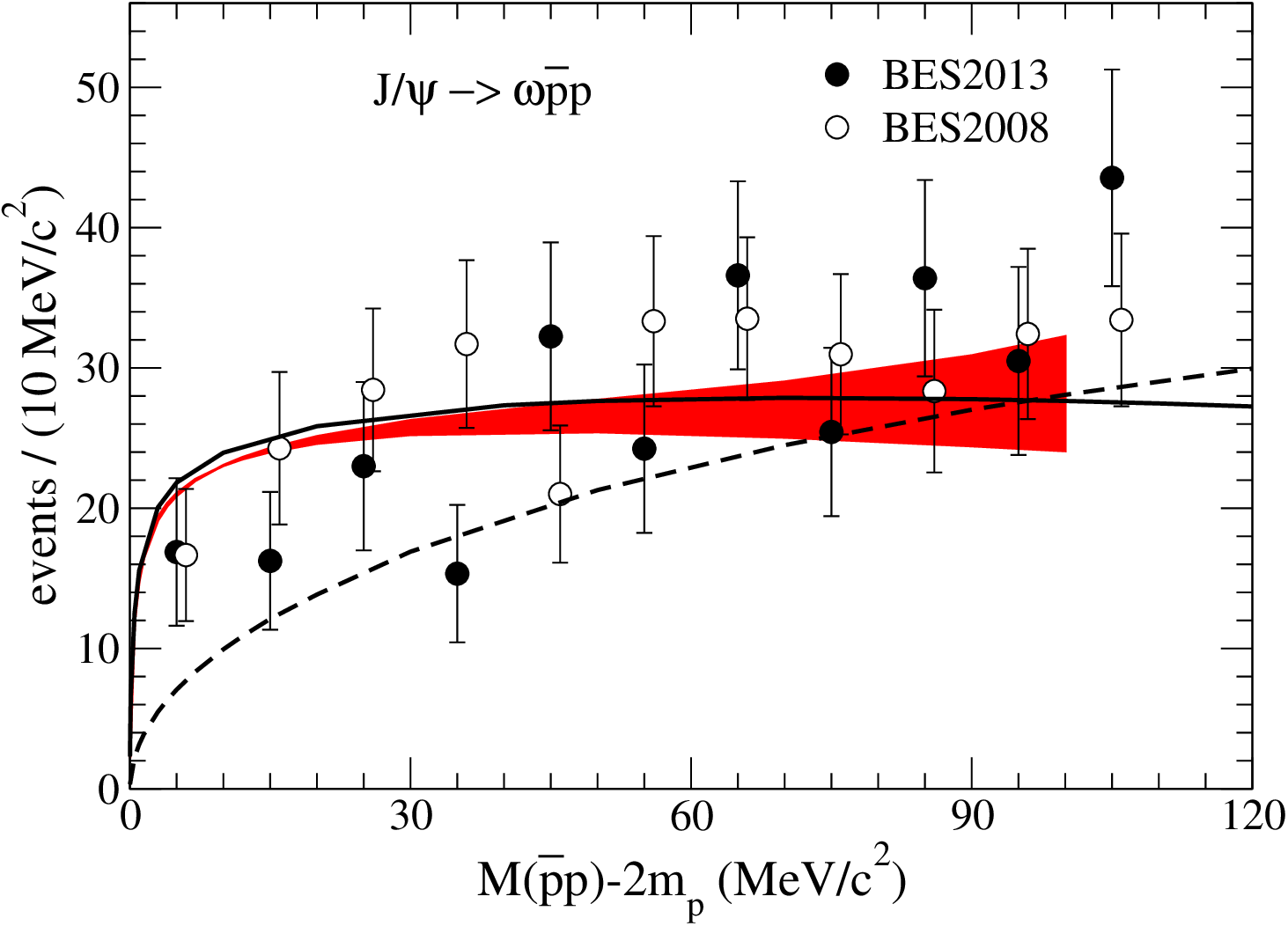}
\caption{$\ppbar$ spectrum for the decay $\jpsi\to\omega\ppbar$. 
Same description of curves as in Fig.~\ref{fig:Jpsipi}. 
Data are taken from Refs.~\cite{BESJpsiomega2008,BESJpsiomega2013}.
The measurement of Ref.~\cite{BESJpsiomega2013} is adopted for the scale. 
The data for the measurement from 2008 have been shifted slightly to the right, cf. text.
}
\label{fig:Jpsiomega}
\end{center}
\end{figure}

Let us now look at the decay $\jpsi\to\omega\ppbar$. In this case the $\ppbar$ state
is produced in the $^1S_0$ partial wave with isospin $I=0$, cf. Table~\ref{tab:JPC}. 
Our results based on the J\"ulich model A(OBE) \cite{modelA1} (solid curve) and 
the chiral potential constructed in Ref.~\cite{JHEP} (band) are shown in Fig.~\ref{fig:Jpsiomega} 
and compared to data from the BES Collaboration \cite{BESJpsiomega2008,BESJpsiomega2013}.
As can be seen from Fig.~\ref{fig:Jpsiomega}, the predictions agree rather well with the 
measured $\ppbar$ invariant mass spectrum in the energy range considered. Also here 
differences between the results based on the chiral potential and A(OBE) are small. 
Actually, it seems that for this particular $\NNbar$ partial wave there is no 
strong dependence on the employed FSI formalism. Our results for A(OBE) based on
the Migdal-Watson approximation, published in \cite{Haidenbauer1}, are qualitatively
very similar to the ones we get now within the DWBA. 

\subsection{Radiative decays}

In the $\jpsi\to\gamma\ppbar$ and $\psip\to\gamma\ppbar$ decays the isospin is no 
longer conserved and, in principle, the final $\ppbar$ state can have any admixture
of the isospin 0 and 1 components. In our previous works the $I=1$ amplitude
was used for $\jpsi\to\gamma\ppbar$ \cite{SibirtsevPRD} while for $\psip\to\gamma\ppbar$ 
the isospin averaged amplitude, $T^{\ppbar} = (T^{I=0} + T^{I=1})/2$, was found
to yield a good agreement with the measurements \cite{Haidenbauer3}. 
Results based on an $\NNbar$ interaction derived from the quark model, presented in
Ref.~\cite{Entem}, suggest that the FSI effects of both isospin components might
be roughly in line with the data, while apparently in 
Refs.~\cite{Loiseau1,Loiseau2,Loiseau3} only the isospin $0$ amplitude was considered.
The BES Collaboration argues in favor of a decay into a pure $I=0$ $\ppbar$ state, guided 
by the experimental observation that apparently $I=1$ states are suppressed in $\jpsi$ 
radiative decays \cite{Ablikim:X1835}. Indeed, the branching fraction of $\jpsi\to\gamma \pi^0$ 
is very small as compared to $\jpsi\to\gamma \eta$ \cite{PDG}. But one must also 
say that there are only a few candidates listed in \cite{PDG} for a decay of 
$\jpsi$ into $\gamma$ and a pure $I=1$ hadronic channel. And, in case of 
$\jpsi \to\gamma\rho\omega$ for example, only an upper limit of the branching fraction 
is known. 

Note that for the reaction $\jpsi\to\gamma\ppbar$ a partial-wave analysis has
been performed \cite{gamma2012}. It suggests that the near-threshold enhancement is 
dominantly in the $J^{PC} = 0^{-+}$ state, which means that the $p{\bar p}$
system should be in the $^{1}S_0$ partial wave.

As already shown above, using the J\"ulich model A(OBE) as input, the mass dependence 
of the near-threshold $\ppbar$ spectrum (and specifically the pronounced peak) is no 
longer reproduced when the refined treatment of the FSI is employed. It turns out that
the same is also the case for the chiral EFT potential of Ref.~\cite{JHEP}.
 
In the present study we adhere to the hypothesis that the enhancement in the $\gamma\ppbar$ 
channel is connected with the $\ppbar$ FSI. Then, there are two options: 
First, we can dismiss the assumption
that the produced $\ppbar$ state consists only of the $I=1$ component alone (made 
in our earlier work \cite{SibirtsevPRD} and also in the calculation based on the EFT
interaction mentioned right above) and allow for an arbitrary mixture of the $I=0$ 
and $1$ amplitudes. 
Second, we can question the amplitude in the $^{1}S_0$ partial wave as predicted 
by the employed J\"ulich A(OBE) and chiral EFT $\NNbar$ potentials. Since the one produced by the 
latter interaction was fixed by a fit to the partial-wave analysis of Zhou and 
Timmermans \cite{Zhou} this implies that we have to depart from the results 
of that analysis. 

Clearly, for physical (and practical) reasons we still want to stay as close as possible 
to the solution given in Ref.~\cite{Zhou} which reproduces the considered $\NNbar$ 
data very well. Thus, we allow only minimal variations in the $^{1}S_0$ partial 
wave and keep all other partial waves fixed. 
Furthermore, we require that all $\NNbar$ scattering observables in the low-energy
region remain practically unchanged. This concerns the total, elastic ($\ppbar\to\ppbar$), 
and charge-exchange ($\ppbar\to \nnbar$) cross sections, and also the differential cross 
sections. Since at low energies those observables are dominated by the $^3S_1$ partial 
wave and the weight of the $^{1}S_0$ amplitude is fairly small, there is some freedom 
for variations even under such strict constraints. 
 
We will consider only variations in the $^{31}S_0$ partial wave, i.e. in the $I=1$ 
amplitude. The $^{11}S_0$ potential is kept as in Ref.~\cite{JHEP}. Given 
the fact that the $\ppbar$ invariant mass spectrum for $\jpsi\to\omega\ppbar$ is 
well reproduced by the $^{11}S_0$ amplitude we do not see any reasons to introduce 
modifications in this partial wave. Recall that the $\gamma\ppbar$ and 
$\omega\ppbar$ channels involve the very same amplitudes, see the selection rules 
in Table~\ref{tab:JPC}. Thus, the assumption that isospin is conserved in the 
hadronic decay rules out that the strong enhancement seen for $\gamma\ppbar$ 
can be directly associated with FSI effects due to the $I=0$ amplitude. Indeed, 
any appreciable modification of the $I=0$ amplitude would automatically spoil 
the reproduction of the $\omega\ppbar$ data.
Note, however, that, in principle, one cannot exclude that isospin conservation is 
also violated in hadronic decays, see, e.g., Ref.~\cite{Ablikim:iso}.

In the following we examine the two options jointly. We regard two exemplary 
combinations of the two isospin amplitudes, namely the ``standard'' one,
$T = T^{\ppbar} = (T^{0} + T^{1})/2$, and also one with a predominant 
$I=0$ component, $T = (0.7\,T^{0} + 0.3\,T^{1})$. For both cases we then
perform a combined fit to the $\ppbar$ invariant mass spectrum for 
$\jpsi\to\gamma\ppbar$ (up to excess energies of $67.5$ MeV) 
and to the $\NNbar$ partial-wave cross sections of the 
$^{1}S_0$ amplitude as determined in the PWA of Zhou and Timmermans \cite{Zhou}. 
Results for the $\ppbar$ invariant mass spectrum are reported below while 
details and results for the $\NNbar$ sector are summarized in the Appendix. 

\begin{figure}[htbp]
\begin{center}
\includegraphics[height=60mm,clip]{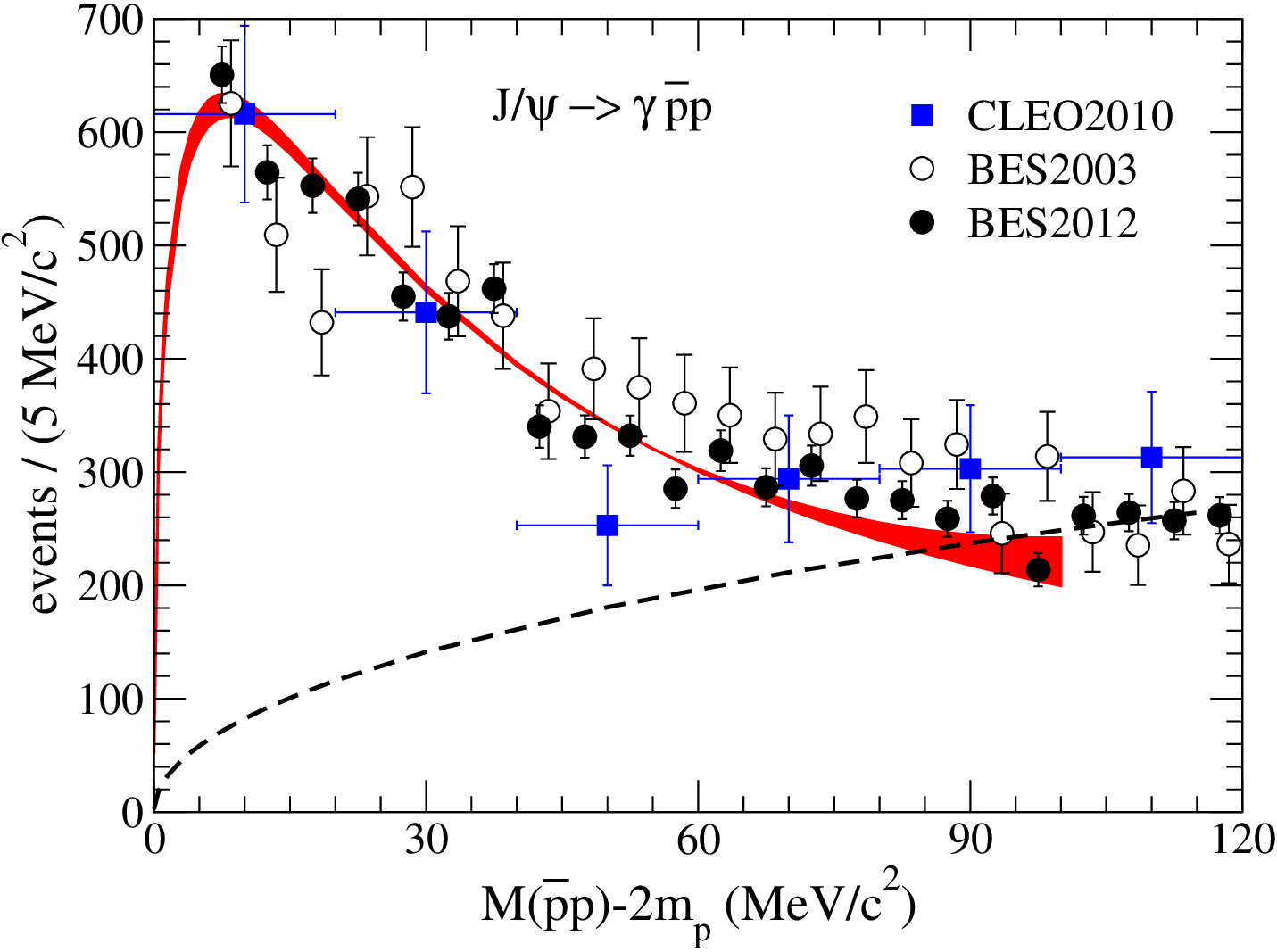}
\caption{$\ppbar$ spectrum for the decay $\jpsi\to\gamma\ppbar$. 
The band represents the result with the refitted chiral NNLO potential, see text. 
The dashed curve denotes the phase space behavior. 
Data are taken from Refs.~\cite{gamma2003,gamma2012,CLEOpsip}.
The measurement of Ref.~\cite{gamma2012} is adopted for the scale. 
The data for the BES measurement from 2003 have been shifted slightly to the right, cf. text.
}
\label{fig:Jpsigamma}
\end{center}
\end{figure}

The decay rate for $\jpsi\to\gamma\ppbar$ based on the refitted $\NNbar$ interaction 
is shown in Fig.~\ref{fig:Jpsigamma}. The results are for the combination
$T = (T^{0} + T^{1})/2$. One can see that now the pronounced peak near $10$ MeV 
is very well described by the FSI. At the same time our (former) $\NNbar$ results 
are also reproduced, c.f. the Appendix. 
Interestingly, the modified potential generates a bound state in the $^{31}S_0$ channel
which was not the case for the interaction presented in Ref.~\cite{JHEP}. 
For example, for the cutoff combination $\{\Lambda,\,\tilde\Lambda\}=$ \{450\,\text{MeV},\,500\,\text{MeV}\} 
the bound state is located at $E_B=(-36.9-{\rm i}\, 47.20)\,\text{MeV}$, where the 
real part denotes the energy with respect to the $\NNbar$ threshold. As it happens, 
this bound state is not very far away from the position of the $X(1835)$ resonance
found by the BES Collaboration in the reaction $\jpsi\to\gamma\pi^+\pi^-\eta'$
\cite{Ablikim:X1835,Ablikim:X1835a}. 
That resonance was interpreted as a possible signal for a $\NNbar$ bound states in
several investigations. But, be aware, our bound state is in the $I=1$ channel {\it and not} 
in $I=0$ as advocated in publications of the BES Collaboration \cite{Ablikim:X1835} 
and of other authors \cite{Loiseau2}. 
In any case, we want to stress that the actual value we 
get for the binding energy should be viewed with caution. 
As mentioned, we examined also the combination $T=0.7\,T^{0}+0.3\,T^{1}$, 
and with it we can achieve likewise a simultaneous description of the 
$\jpsi\to\gamma\ppbar$ data and the $\ppbar$ scattering cross section with
similar quality. However, in this case the position of the bound state is 
around $E_B=(-14.8-{\rm i}\, 39.7)\,\text{MeV}$. Clearly, the data above the
$\NNbar$ threshold do not allow to determine the binding energy reliably
given that the bound state might be $30$ or $40$ MeV below the threshold
and has a sizable width. 

Note that we do not show in Fig.~\ref{fig:Jpsigamma} the data points in the lowest bin 
from the BES experiments. For energies below 5 MeV the Coulomb interaction has a significant 
influence and likewise the difference between the $\ppbar$ and $\nnbar$ thesholds
plays a role. Both effects are not included in the present calculation. Indeed, 
because of the strong energy dependence very near threshold, one would need to 
take into account also the finite momentum resolution of the experiment for a 
sensible comparison with the data. 

\begin{figure}[htbp]
\begin{center}
\includegraphics[height=60mm,clip]{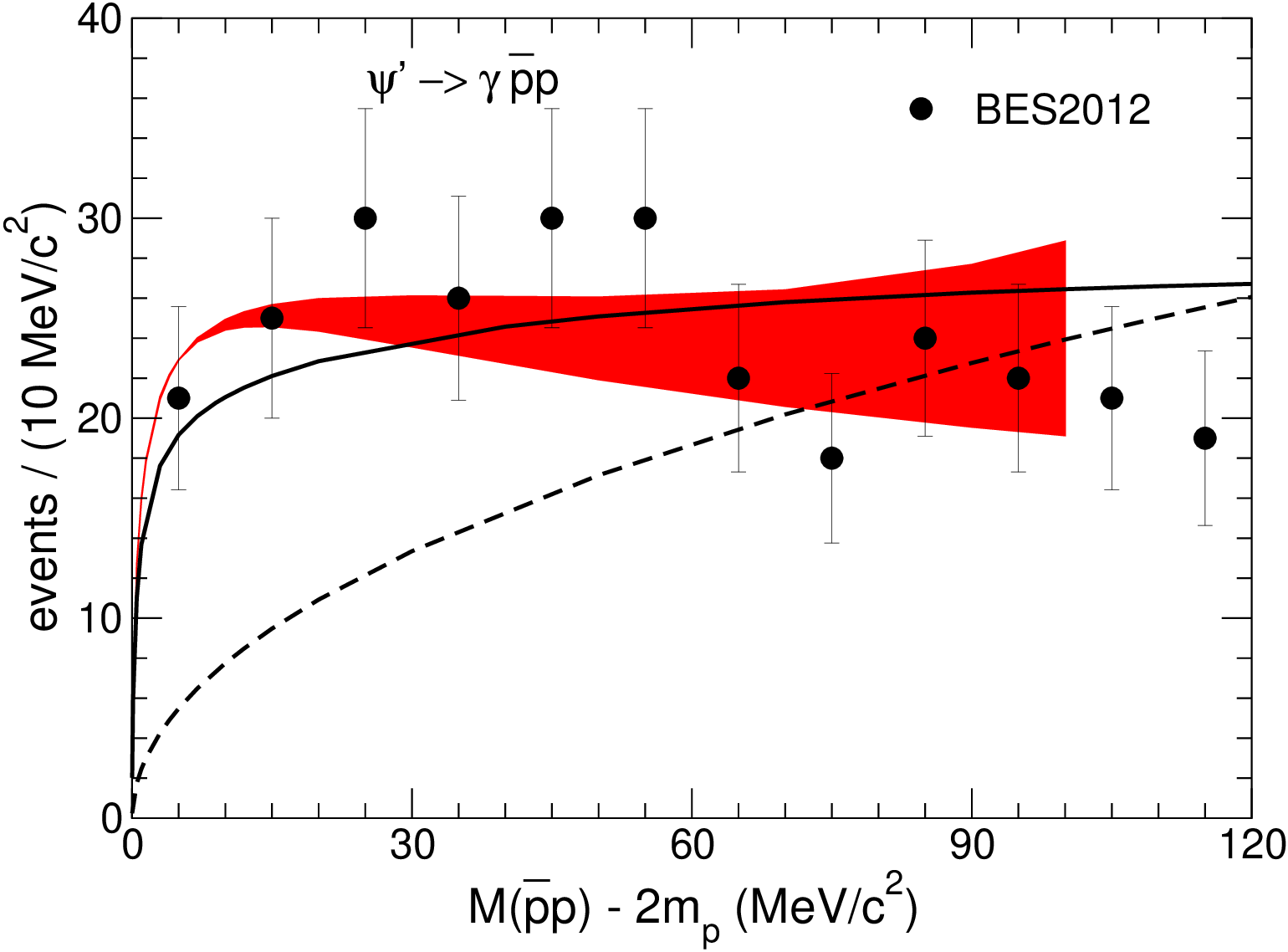}
\quad 
\includegraphics[height=60mm,clip]{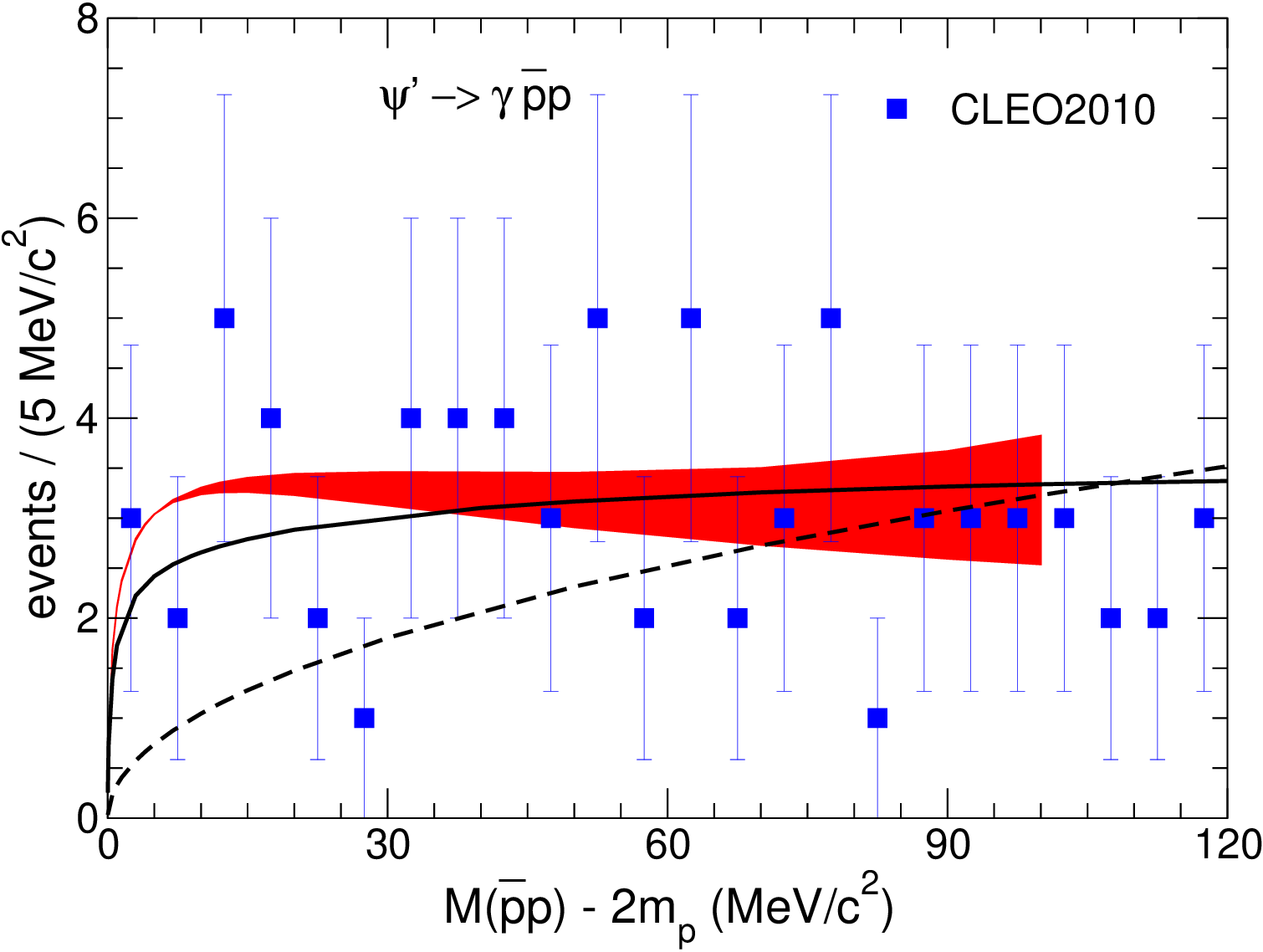}
\caption{$\ppbar$ spectrum for the decay $\psip\to\gamma\ppbar$. 
The band represents the result with the refitted chiral NNLO potential, see text. 
The solid curve is the result for the $\NNbar$ interaction A(OBE) \cite{modelA1}.  
The dashed curve denotes the phase space behavior. 
Data are taken from Refs.~\cite{gamma2012,CLEOpsip}.
}
\label{fig:psipgamma}
\end{center}
\end{figure}

There are also experimental results for the decay $\psip\to\gamma\ppbar$ \cite{gamma2012,CLEOpsip}.   
While the statistics is not as high as for the $\jpsi\to\gamma\ppbar$ case, nonetheless, 
the recent data 
from the BES Collaboration \cite{gamma2012} provide clear evidence that, contrary to 
$\jpsi\to\gamma\ppbar$, in this channel there is no prominent near-threshold peak, but 
still a significant enhancement 
as compared to the pure phase-space behavior, see Fig.~\ref{fig:psipgamma}. This is 
interesting because the quantum numbers of the particles involved in the two reactions
are identical and, therefore, one would expect naively to see similar effects from the
$\ppbar$ FSI. However, in the $\psip\to\gamma\ppbar$ decay isospin is likewise not conserved 
and, in particular, the reaction amplitude can have a different admixture of the isospin-$0$ 
and isospin-$1$ components. Indeed, when we assume, for example, that for $\psip\to\gamma\ppbar$
the final $\ppbar$ state is given by the combination $T=0.9\,T^{0}+0.1\,T^{1}$ we can 
describe the $\ppbar$ invariant mass spectrum measured in this reaction very well, 
as demonstrated in Fig.~\ref{fig:psipgamma}. But a somewhat smaller or larger admixture ($\pm$ 5-10~\%)
of the isospin-$0$ component would still yield results that are compatible 
with the data. 
Note also that the isospin-$1$ $T$-matrix from the refitted $^{31}S_0$ potential is 
employed here, i.e. the same amplitude as in our $\jpsi\to\gamma\ppbar$ calculation. 

Results based on the $\NNbar$ model A(OBE) are also shown in Fig.~\ref{fig:psipgamma} 
(solid lines). Here agreement is found for the isospin combination $T=(T^{0}+T^{1})/2$.

\begin{figure}[htbp]
\begin{center}
\includegraphics[height=60mm,clip]{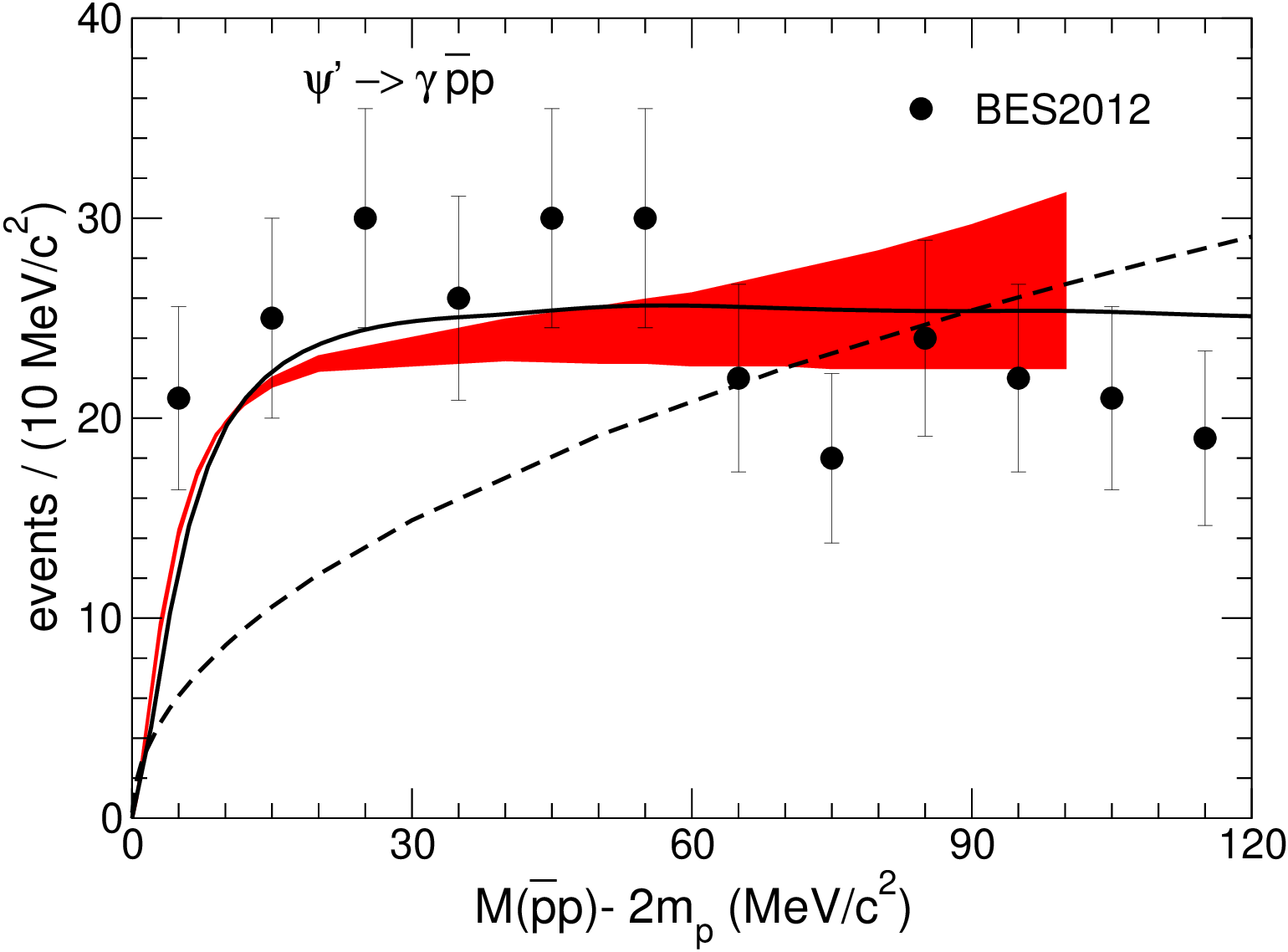}
\caption{$\ppbar$ spectrum for the decay $\psip\to\gamma\ppbar$. Same description of curves as in 
Fig.~\ref{fig:psipgamma}, however, the $^{13}P_0$ partial wave is used for generating  
the $\NNbar$ FSI effects.
Data are taken from Ref.~\cite{gamma2012}.
}
\label{fig:psipgamma13P0}
\end{center}
\end{figure}

The branching ratios of $\psip\to\gamma\chi_{cJ}$ ($J=0, 1, 2$) are around 10\%, for each of the 
$\chi_{cJ}$'s \cite{PDG}. Together they amount to about 30\%, which is orders of magnitude 
larger than all other radiative decay modes. 
Thus, it is quite possible that in the radiative decay of $\psip$ the $\ppbar$ 
pair is produced predominantly via one of the $\chi_{cJ}$ resonances acting as doorway state. 
If so, then the $\ppbar$ state must emerge in a $P$-wave, see Tab.~\ref{tab:JPC}. 
Therefore, we performed also calculations where we explored such a scenario. It turned out that 
the assumption of a transition via the $\chi_{c0}$ resonance which then leads to a $\ppbar$ 
final state in the $^3P_0$ partial wave yields results that agree fairly well with the data. 
The corresponding event distribution for the final $\ppbar$ pair is presented in 
Fig.~\ref{fig:psipgamma13P0} where the isospin-$0$ amplitude predicted by the considered
$\NNbar$ interactions was employed.
Anyway, the masses of the $\chi_{cJ}$, $J=0, 1, 2$ states are $3415$, $3511$, and $3556$~MeV, 
respectively \cite{PDG}. Thus, the $\NNbar$ theshold is very far away from the nominal masses of 
those resonances and, therefore, only the very tail of the $\chi_{cJ}$'s can contribute to the
$\ppbar$ spectrum at those low invariant masses considered in our investigation.  

\subsection{Discussion}
The scenario outlined above allows us to describe consistenly (and quantitatively) the 
near-threshold enhancement seen in the $\ppbar$ invariant mass spectrum of various $\jpsi$ 
and $\psip$ decays in terms of FSI effects. 
In particular, we can reproduce the moderate enhancement seen in the reactions 
$\jpsi\to \omega\ppbar$ and $\psip\to \gamma\ppbar$ as well as the rather large 
enhancement in the $\jpsi\to \gamma\ppbar$ channel. The analysis of the latter indicates 
the possible existence of a $\NNbar$ bound state. 
However, contrary to the suggestion of the BES Collaboration \cite{Ablikim:X1835} and the 
theoretical studies of the Paris group \cite{Loiseau2}, this bound state would be in the 
isospin-$1$ channel and not in isospin 0!
Therefore, in the following, let us discuss our scenario and possible alternatives in detail. 

Near the $\ppbar$ threshold the reactions $\jpsi\to \gamma\ppbar$, $\psip\to \gamma\ppbar$, 
and $\jpsi\to \omega\ppbar$ are all governed by the same $\NNbar$ partial wave, namely the
$^1S_0$ (cf. Table~\ref{tab:JPC}). The assumption that isospin is conserved in the
hadronic decay $\jpsi\to \omega\ppbar$, together with the observed moderate enhancement
in the pertinent $\ppbar$ invariant mass spectrum, practically excludes that the exceptionally
large enhancement in the $\jpsi\to \gamma\ppbar$ decay has anything to do with the
isospin-$0$ $\NNbar$ amplitude. Actually, as shown in our analysis, the two measurements can be 
only reconciled if we assume that the decay into $\gamma\ppbar$ involves a substantial isospin-$1$ 
amplitude. Of course, it could be possible that there is a strong violation of isospin conservation in 
the hadronic decay $\jpsi\to \omega\ppbar$. However, we believe that this is much less likely 
than a sizable isospin-$1$ admixture in the radiative reaction $\jpsi\to \gamma\ppbar$ where isospin 
is not conserved anyway. Another option would be that the decay $\jpsi \to \omega\ppbar$ 
leads predominantly to $\NNbar$ $P$-waves -- even close to threshold -- and only the 
reaction $\jpsi\to \gamma\ppbar$ is dominated by the decay into the $^1S_0$ partial wave. 
While a dominance of $P$-waves might be indeed plausible for $\psip\to \gamma\ppbar$, 
as discussed above, at the moment there is no experimental evidence that it could be also the 
case for the $\omega$ channel. 
Clearly, here measurements of the angular distributions for the $\omega\ppbar$ case, analogous to
those available for $\gamma\ppbar$ \cite{gamma2012}, would be very useful. 

\begin{figure}[htbp]
\begin{center}
\includegraphics[height=60mm,clip]{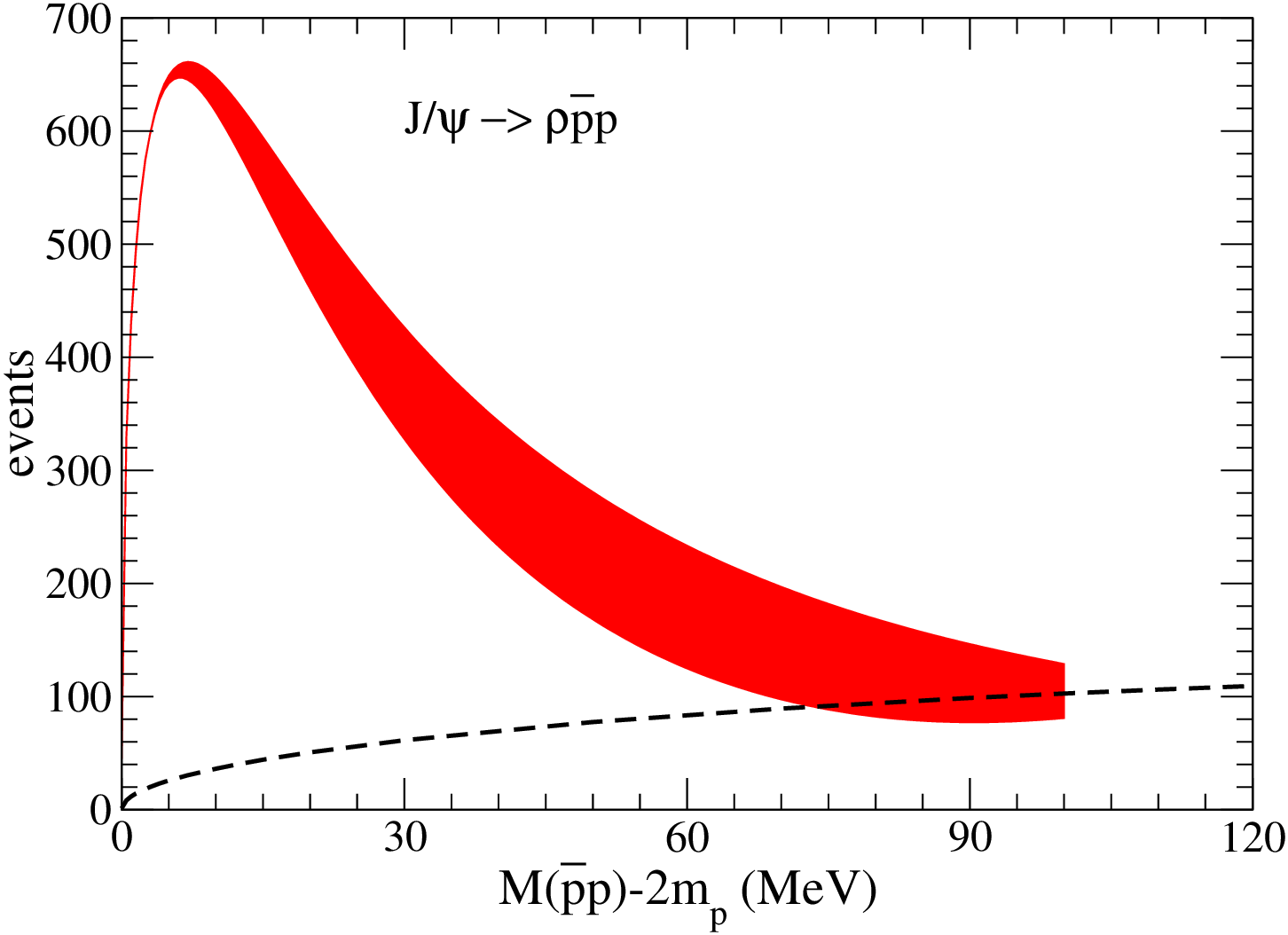}
\caption{$\ppbar$ spectrum for the decay $\jpsi\to\rho\,\ppbar$.
The band represents the result with the refitted chiral NNLO potential, see text. 
The dashed curve denotes the phase space behavior. 
}
\label{fig:Jpsitho}
\end{center}
\end{figure}

What if a genuine resonance is responsible for the enhancement observed in the decay 
$\jpsi\to \gamma\ppbar$? Of course, such a resonance should not couple strongly to the $\NNbar$ 
channel, because otherwise it will contribute significantly to the (direct) $\NNbar$ interaction. 
Then, in turn, it would contribute to the $\NNbar$ FSI effects in the pertinent channel, i.e. it 
should be also seen in $\omega\ppbar$, for example. 
A resonance that couples strongly
to $\jpsi$ and only rather weakly to $\NNbar$ should be seen in other $\jpsi$ decay channels.
In principle, the $X(1835)$ found by the BES Collaboration in the reaction
$\jpsi\to \gamma\pi^+\pi^-\eta'$ \cite{Ablikim:X1835,Ablikim:X1835a} could be a 
candidate for such a resonance. But then we expect it to be absent in the corresponding 
reaction $\jpsi\to \omega\pi^+\pi^-\eta'$, say -- otherwise one would again have difficulties 
to explain simultaneously the rather moderate enhancement for the $\omega\ppbar$ channel.
Indeed, it would be interesting to investigate the latter $\jpsi$ decay channel experimentally. 

In any case, the scenario favored by us where the exceptionally strong near-threshold enhancement 
in the reaction $\jpsi\to\gamma\ppbar$ is primarily due to strong FSI effects in the $^1S_0$ 
$\NNbar$ amplitude with isospin $I=1$ can be tested experimentally. If this scenario is
correct then one should see a similarly strong enhancement in other decay channels where 
near threshold the $\NNbar$ system is produced in the same partial wave. This
applies first of all to the reaction $\jpsi\to\rho\,\ppbar$ where the $\NNbar$ state has 
to have $I=1$, provided that isospin is conserved in this strong decay. We present our
predictions for the corresponding invariant mass spectrum in Fig.~\ref{fig:Jpsitho}. 

A measurement of $\chi_{c0}$ decaying into $\pi^- p\bar{n}$ would be also rather
interesting. In this case, near threshold the $p\bar{n}$ state is likewise produced in 
the $^1S_0$ partial wave and, moreover, it has to be in isospin $I=1$, see Table~\ref{tab:JPC}.
Data reported in Ref.~\cite{Ablikim:chi} suggest that there is a large enhancement
in the $p\bar{n}$ invariant mass spectrum in the low-energy region. However, the
invariant-mass resolution is still fairly poor and does not allow for any reliable
conclusions.

\section{Summary}

In the present paper we have provided a systematic analysis of the near-threshold enhancement 
in the $\ppbar$ invariant mass spectrum, as observed in various experiments of the
decay reactions $\jpsi \to x p\bar p$ and $\psip(3686) \to x p\bar p$, with $x = \gamma,\, \omega,\, 
\pi,\, \eta$. The enhancement is assumed to be due to the $\NNbar$ final-state 
interaction (FSI) and the pertinent FSI effects are evaluated in an approach that is 
based on the distorted-wave Born approximation. 
For the $\NNbar$ interaction a potential derived within chiral effective
field theory and fitted to results of a recent partial-wave analysis of 
$\ppbar$ scattering data \cite{Zhou} is employed. For comparison, a phenomenological model 
constructed by the J\"ulich group and used by us in earlier studies of $\jpsi$ and $\psip$
decays is also utilized. 
It is found that the near-threshold spectrum of all considered decay reactions
can be reproduced simultaneously and consistently by our treatment of the $\ppbar$ FSI. 
Specifically, the moderate enhancement seen for $\pi^0 \ppbar$, $\eta \ppbar$, and
$\omega \ppbar$ final states is well described by the $\NNbar$ interaction in the
relevant $^3S_1$ and $^1S_0$ partial waves as determined in the partial-wave analysis.

The situation is more complicated for the process $\jpsi\to\gamma\ppbar$ where
there is a rather large near-threshold enhancement. While the pertinent $\ppbar$ invariant
mass spectrum was reproduced in our previous work \cite{SibirtsevPRD} that was based 
on the Migdal-Watson approach, this is no longer the case for the more realistic
treatment of FSI effects employed in the present study. 
However, we can show that a modest modification of the interaction in the $I=1$ $^{1}S_0$ $\NNbar$ 
channel -- subject to the constraint that the corresponding partial-wave cross sections
for $\ppbar \to \ppbar$ and $\ppbar \to \nnbar$ remain practically unchanged at low energies -- 
allows one to reproduce the events distribution of the radiative $\jpsi$ decay, 
and consistently all other decays. In this context the decay $\jpsi\to\omega\ppbar$ plays 
a crucial role. The moderate enhancement observed in this channel, together with the
fact that the produced $\ppbar$ system has to be in $I=0$ (assuming that isospin
is conserved in this purely hadronic decay) implies that the strong variation seen in
the $\gamma\ppbar$ case has to come primarily from the $I=1$ $^{1}S_0$ $\NNbar$ interaction. 

It turns out that the modified $I=1$ $^1S_0$ interaction that can reproduce the 
$\ppbar$ invariant mass spectrum in the reaction $\jpsi\to \gamma\ppbar$
predicts a $\NNbar$ bound state.
Previous investigations suggested that there could be such a bound state, but in the isospin $I=0$
channel \cite{Loiseau2}. Also the BES Collaboration favored an $I=0$ bound state, 
being led by their observation of the $X(1835)$ resonance in the reaction 
$\jpsi\to\gamma\pi^+\pi^-\eta'$ \cite{Ablikim:X1835}. 
Interestingly, the value we get for the binding energy is comparable to the mass of the 
$X(1835)$. However, we want to stress that one should view our value with great caution. 
First, due to the unknown fraction of the $I=0$ and $I=1$ components in the final $\ppbar$ 
state for the radiative decay there is a sizable uncertainty in the actual value.
Moreover, one should be aware that, in general, any data above the reaction threshold, like
the $\ppbar$ invariant mass spectrum in the present case, do not allow to pin down the 
binding energy reliably given that the bound state might be $30$ or $40$ MeV below the 
threshold and has a sizable width. Actually, at this stage we cannot exclude that 
an alternative fit of similar quality to the invariant mass spectrum and to the 
near-threshold $\NNbar$ scattering data is possible without a bound state
in the $I=1$ $^1S_0$ $\NNbar$ partial wave. 

Another interesting implication of our study is that $\ppbar$ invariant mass spectra as 
measured in heavy meson decays could be indeed very useful as further constraint 
for the determination of the $\NNbar$ partial-wave amplitudes, provided that those data 
are of high statistics and high resolution like the ones for $\jpsi\to \gamma\ppbar$. 
This is of specific relevance for the near-threshold region. 
Here the available $\NNbar$ observables are dominated by the $^3S_1$ partial wave
whereas the weight of the $^{1}S_0$ amplitude is fairly small. At such low energies
direct $\ppbar$ scattering experiments for measuring spin-dependent observables 
that would allow one to disentangle the spin-singlet and triplet 
contributions are rather difficult (if not impossible) to perform. 

\section*{Acknowledgements}
One of the authors (XWK) acknowledges communications with Prof.~Changzheng Yuan 
concerning the status of the various BES experiments.
This work is supported in part by the DFG and the NSFC through
funds provided to the Sino-German CRC 110 ``Symmetries and
the Emergence of Structure in QCD'' and by the EU Integrated
Infrastructure Initiative HadronPhysics3.

\appendix
\section{$\NNbar$ interaction in the $^1S_0$ partial wave}
%\countzero
A comprehensive description of our $\NNbar$ potential derived within chiral EFT can
be found in Ref.~\cite{JHEP}, where all technical details are given. 
Here we focus only on those ingredients that are relevant for the alternative 
description of the $^1S_0$ partial wave with isospin $I=1$ that is employed 
in Sect. III. In this case a refit of the low-energy constants (LECs) in
the contact terms was performed. Up to next-to-next-to-leading order (NNLO)
the corresponding contribution to the potential is
given by \cite{JHEP}
\bea 
\text{Re}\,V(^1S_0)&=&\tilde C_{^1S_0}+ C_{^1S_0}(p^2+p'^2)\nl 
\text{Im}\,V(^1S_0)&=&V_{\text{ann}}(^1S_0)\nl
& =& -i\, (\tilde C_{^1S_0}^a+C_{^1S_0}^ap^2)
(\tilde C_{^1S_0}^a+C_{^1S_0}^ap'^2), \nl
\label{ANN} 
\eea
where $p \, (p')$ is the modulus of the three-momentum for the initial (final) state in the 
center-of-mass system (CMS). In Ref.~\cite{JHEP}, the values for these LECs
$\tilde C_{^1S_0}, \cdots, C^a_{^1S_0}$ were obtained by fitting to the results
of the partial-wave analysis (PWA) for this particular partial wave provided 
in Ref.~\cite{Zhou}.

\begin{figure}[htbp]
\begin{center}
\vskip 0.3cm
\includegraphics[scale=0.50,clip=true]{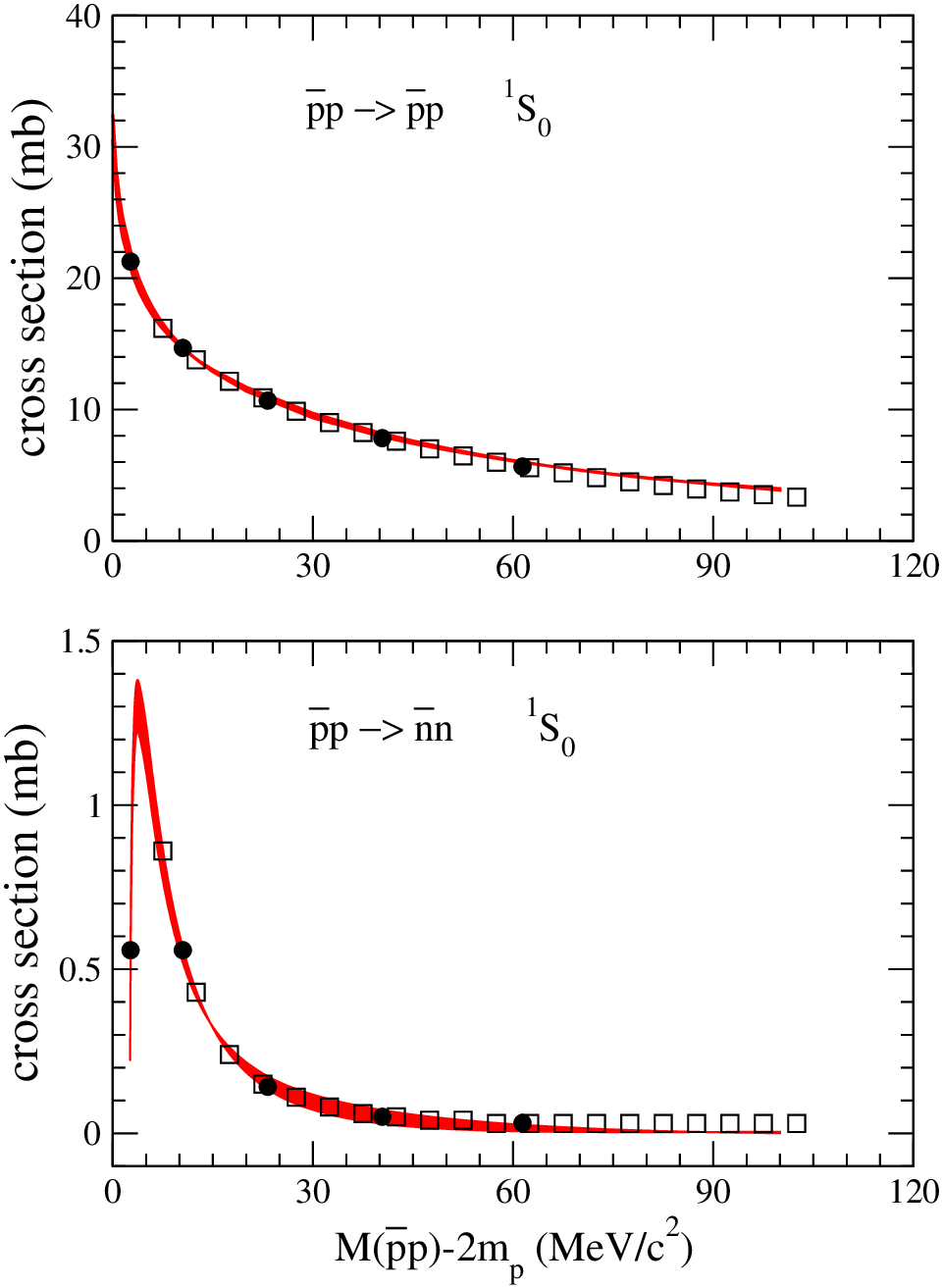}
\caption{The $^1S_0$ partial-wave cross sections as a function of the excess energy. 
The squares represent the results for the published NNLO potential \cite{JHEP} with
the cutoff combination \{450\,\text{MeV}, 500\,\text{MeV}\}.
The circles indicate the cross sections for the partial-wave amplitudes of Ref.~\cite{Zhou}.
The bands show the results based on the refitted isospin-$1$ $^1S_0$ amplitudes.
}
\label{fig:pwcs}
\end{center}
\end{figure}

Now the LECs appearing in the $^{31}S_0$ potential in Eq.~(\ref{ANN}) are fitted to both,
the $\NNbar$ $^{31}S_0$ partial-wave cross section up to laboratory energies of $125$ MeV
and the $\jpsi\to\gamma\ppbar$ event distribution (up to excess energies of $67.5$~MeV).
With regard to the partial-wave cross section we fit to the one produced by the original 
NNLO interaction of Ref.~\cite{JHEP}. 
This makes sure that we stay also as close as possible to the results of the PWA 
from Ref.~\cite{Zhou}. As can be seen in Ref.~\cite{JHEP} the phase shifts and 
inelasticities of our EFT potential are basically identical to the ones from the
PWA up to laboratory energies of around 125 MeV. 
The LECs resulting from the fitting procedure are listed in Table~\ref{tab:LEC}. 
In this context we want to mention that there is a mistake in Tables 1 and 2 of 
Ref.~\cite{JHEP}, i.e. in the list of the LECs for our original NLO and NNLO interactions. 
In case of the $^1S_0$ partial wave the parameters for $C_{^1S_0}$ and $\tilde C^a_{^1S_0}$ 
are mixed up (for both isospins). For example, this means that the parameters given 
in the 2nd line are those for $\tilde C^a_{^1S_0}$ and the ones in the 3rd line 
are those for $C_{^1S_0}$. 

Let us first look at the scattering length and compare the present one with that
of the original interaction \cite{JHEP}, cf. Table~\ref{tab:levelshifts}.
The corresponding level shifts and widths for the antiproton hydrogen in the state 
of $1\,^1S_0$ are also compiled in that table.
From these numbers we see that the predictions with the modified $^{31}S_0$ potential 
agree with the original ones within the uncertainty induced by the cutoff variation.
We provide here also some experimental information on these quantities 
\cite{Augsburger:1999yt,Ziegler:1988bp}, though we want to stress that additional 
assumptions have to be made in order to deduce the splitting of the $1$$^1S_0$ level shift 
from the experiment \cite{Gotta,GottaP}.
 
The resulting $^1S_0$ partial cross sections for the reactions $\ppbar\to\ppbar$ and 
$\ppbar\to \nnbar$ are displayed in Fig.~\ref{fig:pwcs}. Here the squares 
represent the results for the published NNLO potential \cite{JHEP} with the cutoff
\{450\,\text{MeV}, 500\,\text{MeV}\} while the bands show our calculation 
with the refitted isospin-$1$ $^1S_0$ amplitude. We see that the latter reproduces 
the former results very well. The circles are the partial-wvae cross sections for
the PWA of Ref.~\cite{Zhou}.

\begin{figure*}
\begin{center}
\includegraphics[height=100mm,clip]{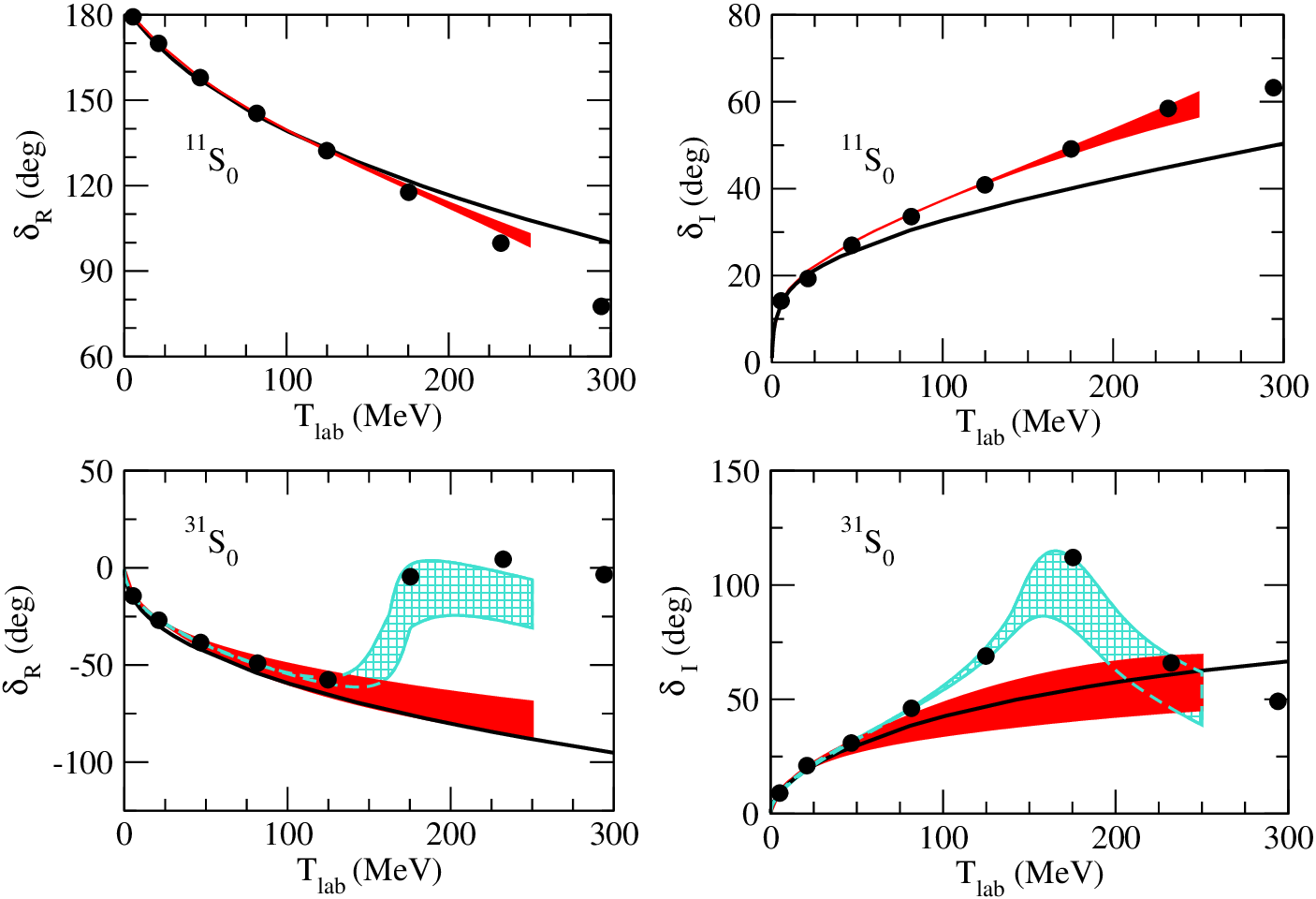}
\caption{Real and imaginary parts of the phase shift in the $^1S_0$ partial waves.
The filled bands show results for the employed EFT interaction up to NNLO. 
The isospin-$1$ phase shifts for the original NNLO potential \cite{JHEP} 
are indicated by the hatched band. The solid line is the result for A(OBE). 
The circles represent the solution of the partial-wave analysis of Ref.~\cite{Zhou}.
}
\label{fig:1S0}
\end{center}
\end{figure*}

Finally, in Fig.~\ref{fig:1S0} we present phase shifts for the $^1S_0$
partial wave. Here the results from the refit are shown by a filled band while
those of the published NNLO potential \cite{JHEP} are indicated by the hatched
band. For convenience we reproduce here also the results for the isospin $0$ 
case from \cite{JHEP} and those of the employed J\"ulich $\NNbar$ potential.

\begin{table}[t]
\begin{center}
\begin{tabular}{||cc||r|r|r|r||}
\hline \hline
& &  &  &  &  \\[-2.5ex]
 \multicolumn{2}{||c||}{LEC}   &  $\{ 450, \; 500\}$   &  $\{ 650, \; 500\}$
&  $\{ 450, \; 700\}$  &  $\{ 650, \; 700\}$ \\[0.5ex]
\hline \hline
& &  &  &  &  \\[-3ex]
\multirow{4}{*}{$I=1$}
& $\tilde C_{ ^1S_0}$   & $0.111$   & $0.035$   & $0.096$  &  $0.005$ \\[0.2ex]
& $C_{^1S_0}$           & $0.080$   & $1.273$   & $0.729$  &  $2.022$ \\[0.2ex]
& $\tilde C^a_{ ^1S_0}$ & $-0.263$  & $-0.204$  & $-0.288$ &  $-0.333$ \\[0.2ex]
& $C^a_{ ^1S_0}$        & $4.876$   & $1.541$   & $4.732$  &  $1.935$ \\[1ex]
\hline \hline
\end{tabular}
\end{center}
\caption{Low-energy constants up to NNLO for the different cutoff combinations
$\big\{ \Lambda \, (\mbox{MeV}), \; \tilde \Lambda \, (\mbox{MeV}) \big\}$.
The values of the $\tilde C_i$ are in units of $10^4$ GeV$^{-2}$ and the $C_i$ in $10^4$ GeV$^{-4}$.
The parameters related to annihilation, $\tilde C^a_i$ and $C^a_i$
(see Eq.~(\ref{ANN})) are in units of $10^2$ GeV$^{-1}$ and $10^2$ GeV$^{-3}$, respectively.
\label{tab:LEC}}
\end{table}

%%% A and Energy shifts
\begin{table*}[htb]
\begin{center}
\begin{tabular}{cccc}
\hline \hline
  {}  & {}  & {} & {}\\[-1.5ex]
  {} & present work & Ref.~\cite{JHEP} & Experiment \\[1ex]
\hline\hline
  {}  & {}  & {} & {}\\[-1.5ex]
$a^{I=0}$ (fm) & \multicolumn{2}{c}{$-0.21-\rm{i}\,(1.21 \cdots 1.22)$} & \\[1ex]
$\, a^{I=1}$ (fm) \, & \,
$(0.97 \cdots 1.07)-\rm{i}\,(0.63 \cdots 0.70)$ \, & \,  
$(1.02 \cdots 1.04)-\rm{i}\,(0.57 \cdots 0.61)$ \, & \\[1ex] 
\hline
$\Delta E$ (eV) & \multirow{2}{*}{$-$(329 $\cdots$ 376)} & \multirow{2}{*}{$-$(302 $\cdots$ 361)} & $-$740 $\pm$ 150 \cite{Ziegler:1988bp} \\[1ex]
& & & $-$440 $\pm$ 75 \cite{Augsburger:1999yt} \\[1ex]
\hline
$\Gamma$ (eV)   &\multirow{2}{*}{(1596 $\cdots$ 1659)}  & \multirow{2}{*}{(1545 $\cdots$ 1589)}  & 1600 $\pm$ 400 \cite{Ziegler:1988bp} \\[1ex]
& & & 1200 $\pm$ 250 \cite{Augsburger:1999yt} \\[1ex]
\hline
\end{tabular}
\end{center}
\caption{$^{1}S_0$ scattering lengths $a$ and hadronic shifts and broadenings in hyperfine 
states of $\bar p$H for $1\,^1S_0$. Results based on the refitted $^{31}S_0$ LECs are given 
and compared with the ones given in Ref.~\cite{JHEP} and with empirical information.
The $^{11}S_0$ scattering length is taken over from Ref.~\cite{JHEP}. 
}
\label{tab:levelshifts}
\end{table*}

\vfill \eject
%%%%%%%%%%%%%%%%%%%%%%%%%%%%%%%%%%%%%%%%%%%%%%%%%%%%

\end{document}